\documentclass[preprint]{aastex}

\usepackage{graphicx}
\usepackage{array}
\usepackage{longtable}
\usepackage{natbib}
\usepackage{amssymb}
\begin{document}
\title{Finding $\eta$ Car Analogs in Nearby Galaxies Using \textit{Spitzer}: \\ I. Candidate Selection}
\author{Rubab~Khan\altaffilmark{1},
K.~Z.~Stanek\altaffilmark{1,2},
C.~S.~Kochanek\altaffilmark{1,2}
}

\altaffiltext{1}{Dept.\ of Astronomy, The Ohio State University, 140
W.\ 18th Ave., Columbus, OH 43210; khan, kstanek, ckochanek@astronomy.ohio-state.edu}

\altaffiltext{2}{Center for Cosmology and AstroParticle Physics, 
The Ohio State University, 191 W.\ Woodruff Ave., Columbus, OH 43210}

\shorttitle{Finding $\Eta$ Car}

\shortauthors{Khan et al.~2011}
 
\begin{abstract}
\label{sec:abstract}

The late-stage evolution of the most massive stars such as $\eta$\,Carinae is controlled 
by the effects of mass loss, which may be dominated by poorly understood eruptive 
mass ejections. Understanding this population is challenging because no true 
analogs of $\eta$\,Car have been clearly identified in the Milky Way or other galaxies. We utilize 
\textit{Spitzer} IRAC images of 7 nearby ($\lesssim4$\,Mpc) galaxies to search for such analogs. 
We find 34 candidates with a flat or rising mid-IR spectral energy 
distributions towards longer mid-infrared wavelengths that emit $>10^5$\,L$_\odot$ in the IRAC bands (3.6 to 8.0\,$\micron$) and are 
not known to be background sources. Based on our estimates for the expected 
number of background sources, we expect that follow-up observations will show that most of 
these candidates are not dust enshrouded massive stars, with an expectation of only $6\pm6$ 
surviving candidates. Since we would detect true analogs of $\eta$\,Car for roughly 
200 years post-eruption, this implies that the rate of eruptions like $\eta$\,Car is 
less than the ccSN rate. It is possible, however, that every M$>40$\,M$_\odot$ 
star undergoes such eruptions given our initial results. In Paper\,II we will characterize the candidates through further analysis 
and follow-up observations, and there is no barrier to increasing the galaxy sample 
by an order of magnitude. 
\end{abstract} 
\keywords{stars: evolution, mass-loss, winds, outflows
--- stars: individual: Eta Carinae
--- galaxies: individual (M33, M81, NGC247, NGC300, NGC2403, NGC6822, NGC7793)}
\maketitle

\section{Introduction}
\label{sec:introduction}
Despite being very rare, massive stars such as luminous blue variable (LBVs), 
red super giants (RSGs), and Wolf-Rayet stars (WRs) play a pivotal role in 
enriching the interstellar medium (ISM) through mass loss~\citep[e.g.,][]{ref:Maeder_1981}. 
Understanding the evolution of these massive (M$\gtrsim30 $\,M$_\odot$) stars 
is challenging even when mass loss is restricted to continuous winds~\citep[e.g.,][]{ref:Fullerton_2006}, but poorly understood 
impulsive mass ejections are probably an equally important, if not dominant 
mass loss mechanism~\citep{ref:Humphreys_1984,ref:Smith_2006,ref:Kochanek_2011c}.
Mass loss also determines the structure of the star at death and hence the 
observed properties of the final core-collapse supernova (ccSN). 
In addition, there is also evidence that some supernova (SN) progenitors undergo 
major mass ejection events shortly before exploding \citep[e.g.,][]{ref:GalYam_2007,ref:Smith_2008}, 
further altering the properties of the explosion and implying a connection 
between some eruptive mass-loss events and death 
\citep{ref:GalYam_2007,ref:Smith_2007b,ref:Kochanek_2011b,ref:Chevalier_2012}.
In two cases, eruptions were observed shortly before the ccSN: Type\,Ib SN\,2006jc 
was spatially coincident with a bright optical transient that occurred in 2004~\citep{ref:Pastorello_2007}, and 
SN\,2009ip underwent a series of outbursts in 2009, 2010, and 2011 before probably 
exploding as a Type\,IIn SN~\citep[see][]{ref:Mauerhan_2012,ref:Prieto_2012,ref:Pastorello_2012}.
These processes are likely metallicity dependent~\citep{ref:Meynet_1994,ref:Heger_2003} 
and there is good evidence that the SNe requiring a dense circumstellar medium (the hyperluminous Type\,IIn)
predominantly occur in lower metallicity galaxies~\citep[e.g.,][]{ref:Stoll_2011,ref:Neill_2011}.

Traditional studies of these massive stars search for them optically and then characterize them
spectroscopically~\cite[e.g.,][]{ref:Bonanos_2009,ref:Bonanos_2010,ref:Clark_2012}. 
This approach is not ideal for probing the episodes of major mass-loss because of 
dust formation in the ejecta. Dense winds tend to form dust, 
although for hot stars the wind must be dense enough to form a pseudo-photosphere 
in the wind~\citep{ref:Davidson_1987} that shields the dust formation region from the UV emission of the star 
\citep{ref:Kochanek_2011c}. The star will then be heavily obscured by dust for an extended 
period after the eruption (see, e.g., \citealp{ref:Humphreys_1994}). The great eruption of $\eta$\,Car between 1840 and 
1860 is the most famous case of a stellar outburst, ejecting $\sim10 M_{\odot}$ 
material before reappearing as a hot star in the 1950s (see, e.g., 
\citealp{ref:Humphreys_1997}). The ejecta are now seen as a dusty nebula around 
the star absorbing and then reradiating $\sim90\%$ of the light in the mid-IR.
This means that dusty ejecta are a powerful and long-lived signature of eruption. 
The emission from these dusty envelopes peaks 
in the mid-IR with a characteristic red color and a rising or flat 
spectral energy distribution (SED) in the \textit{Spitzer} IRAC~\citep{ref:Fazio_2004} bands.

In the Galaxy, stars with resolved shells of dust emission primarily at 
24\,$\micron$ are easily found \citep{ref:Wachter_2010,ref:Gvaramadze_2010}. 
The advantage of the 24\,$\micron$ band is that it can be used to identify dusty ejecta 
up to $10^3 - 10^4$\,years after its formation. A minority of these objects are very luminous stars 
(L\,$\gtrsim10^{5.5}\,$\,L$_\odot$) with massive ($\sim0.1-10\,$\,M$_\odot$) shells (see summaries by 
\citealp{ref:Humphreys_1994,ref:Humphreys_1999,ref:Smith_2006,ref:Smith_2009,ref:Vink_2009}). 
These include AG\,Car~\citep{ref:Voors_2000}, 
the Pistol Star~\citep{ref:Figer_1999}, G79.29$+$0.46~\citep{ref:Higgs_1994}, 
Wray\,17$-$96~\citep{ref:Egan_2002}, and IRAS\,18576$+$0341~\citep{ref:Ueta_2001}. 
These systems are far older ($\gtrsim10^3$\,years) than $\eta$\,Car, which makes it difficult 
to use the ejecta to probe the rate or mechanism of mass-loss. 
Still, the abundance of Galactic shells implies that the rate of 
$\eta$\,Car-like eruptions is on the order of a modest fraction of the ccSN rate~\citep{ref:Kochanek_2011c}. Their emission peaks 
in the shorter IRAC bands when they are relatively young ($\sim10-100$\,years) 
because, as the ejected material expands, 
the dust becomes cooler and the emission shifts to longer wavelengths~\citep{ref:Kochanek_2012a}. 
It is difficult to quantify searches for such objects in our Galaxy as it is 
difficult to determine the distances to the sources and the survey 
volume because we have to look through the crowded and dusty disk of the Galaxy. 
Surveys of nearby galaxies are both better defined and build larger samples 
of younger systems whose evolution can be studied to better understand the mechanism.

With \textit{Spitzer} it is difficult to use the 24\,$\micron$ observations that 
have proved so successful in the Galaxy because of the poor angular resolution. 
However, we have shown that such surveys can be done with IRAC (3.6--8.0\,$\micron$).
In \cite{ref:Thompson_2009} and \cite{ref:Khan_2010}, we characterized the 
extreme AGB star populations that appear to be the progenitors of the 
SN\,2008S-like transients~\citep{ref:Prieto_2008a,ref:Prieto_2008b} 
using archival IRAC images of nearby galaxies. These studies empirically 
confirmed that these $\sim10^{4.5} $\,L$_\odot$ dusty stars are rare but are also relatively 
easy to identify in IRAC images despite the modest angular resolution. 
Next, we examined all the other bright, red mid-IR sources in M\,33, 
and in \citet{ref:Khan_2011} we discovered Object\,X, the brightest mid-IR star 
in M\,33. Object\,X is a L$_{bol}\sim5\times10^5 $\,L$_\odot$, 
M$\gtrsim 30 M_{\odot}$ evolved star obscured by dust formed during 
mass loss events over the last $\sim1$\,century. Its properties 
are similar to those of the Galactic OH/IR star IRC+10420 
\citep{ref:Humphreys_1997,ref:Blocker_1999,ref:Humphreys_2002}, which has a 
complex dusty circumstellar structure resulting from episodic, low-velocity 
mass ejections. We proposed that Object\,X may emerge from its current ultra-short 
evolutionary phase as a hotter post-RSG star analogous to M\,33\,Var\,A 
\citep{ref:Hubble_1953,ref:Humphreys_2006}. 

While Object\,X is intriguing, it likely underwent a period of enhanced, but 
relatively steady, mass loss from the parent star rather than the short transient 
episode of mass loss usually associated with so called ``supernova impostors''~\citep{ref:VanDyk_2000}. It is also an 
order of magnitude less luminous and several times less massive than $\eta$\,Car, 
one of the most luminous ($L_{bol}\sim5\times10^6 $\,L$_\odot$) and massive ($M\simeq 100-150 M_{\odot}$)
stars known \citep{ref:Humphreys_1994}.
No true analog of $\eta$\,Car in mass, luminosity, energetics, mass lost and age has 
been found (see \citealp{ref:Smith_2011,ref:Kochanek_2012a}). 
Quantifying the population of $\eta$\,Car analogs, or their rarity, in the 
local universe can allow us to investigate the rate of giant eruptions of 
the most massive stars. It may also help us answer open questions about the 
evolution of massive stars such as: (1) the frequency of major mass 
ejection events, (2) the number of events per star, 
(3) whether the frequency depends on the metallicity or other stellar 
properties, and (4) whether there is really any relation between mass 
ejections and the so called ``supernova impostors'' (see \citealp{ref:Smith_2011,ref:Kochanek_2012a}).

Here we carry out a pilot study of 7 nearby galaxies within ($\lesssim4$\,Mpc) 
to search for analogs of $\eta$\,Car. We concentrate on galaxies with recent star formation, as 
only these will have large numbers of the short-lived, very massive 
stars that we want to study, but we also include one small, low-mass galaxy 
(NGC\,6822) as a test case (see Table\,\ref{tab:galaxies}).
Section\,\ref{sec:search} describes our methodology for identifying potential $\eta$\,Car analogs in nearby
galaxies using archival \textit{Spitzer} data and sources of contamination. Section\,\ref{sec:candidates} discusses the nature
of the candidates, although a detailed study is deferred to Paper\,II. Section\,\ref{sec:discussion} 
shows how our search method allows us to quantify the selection criteria and to set an interesting limit on the rate of events similar 
to the Great Eruption of $\eta$\,Car in the local universe even before we have completed Paper\,II. 
Finally, in Section\,\ref{sec:conclusions} we outline the future of our approach.
\section{A Search for $\eta$\,Car Analogs}
\label{sec:search}
In this section, we present the methodology of our search for $\eta$\,Car analogs.
First we discuss our data sources and the properties of the targeted galaxies. Next 
we describe the photometry and candidate selection procedures. Then we 
consider contamination due to non-stellar sources. Finally we consider if 
an $\eta$\,Car analog in a nearby galaxy could be hidden in a compact stellar cluster.
\subsection{Targeted Galaxies}
\label{sec:target}
\begin{table*}[t]
\begin{center}
\caption{Properties of Targeted Galaxies}
\label{tab:galaxies}
\begin{tabular}{llcccccc}
\\
\hline 
\hline
\multicolumn{1}{c}{Galaxy} &
\multicolumn{1}{c}{Distance} &
\multicolumn{1}{c}{$M_{B}$} &
\multicolumn{1}{c}{$E(B-V)$} &
\multicolumn{1}{c}{$\log_{10} L(H{\alpha})$} &
\multicolumn{1}{c}{SFR (H${\alpha}$)} &
\multicolumn{1}{c}{Known ccSN} 
\\
\multicolumn{1}{c}{} &
\multicolumn{1}{c}{(Mpc)} &
\multicolumn{2}{c}{} &
\multicolumn{1}{c}{(erg/s)} &
\multicolumn{1}{c}{(M\,$_\odot$/yr)} &
\multicolumn{1}{c}{($<20$\,years)} 
\\
\hline
\hline
NGC\,6822 & 0.46 & $-14.9$ & 0.24  & 39.1 & 0.01  & \dots    \\
M\,33   & 0.96 & $-18.8$ & 0.04  & 40.6 & 0.33 & \dots     \\
NGC\,300 & 1.9 & $-17.7$ & 0.01 & 40.1 & 0.11 & \dots      \\
NGC\,2403 & 3.1 & $-18.7$ & 0.04 & 40.8 & 0.44 & SN\,2004dj (IIP) \\
M\,81   & 3.6 & $-20.1$ & 0.08 & 40.8 & 0.46 & SN\,1993J (IIb) \\
NGC\,247 & 3.6 & $-18.2$ & 0.02 & 40.3 & 0.17  & \dots      \\
NGC\,7793 & 4.1 & $-18.5$ & 0.02 & 40.6 & 0.33 & SN\,2008bk (IIP) \\
\hline
\hline
\end{tabular}
\end{center}
\end{table*}
There are a number of sources of archival {\em Spitzer} data for nearby 
galaxies. In the Local Group, SAGE (Surveying the Agents of a Galaxy's Evolution) and SAGE-SMC 
\citep{ref:Meixner_2006,ref:Gordon_2007} surveyed the LMC and SMC. 
\cite{ref:Barmby_2006} surveyed M\,31 with first results for massive stars 
discussed by \cite{ref:Mould_2008}. M\,33 was observed at 
several epochs which allowed for mid-IR variability studies of M\,33 stars 
\citep{ref:McQuinn_2007,ref:Thompson_2009}.
The \textit{Spitzer} Infrared Nearby Galaxies Survey~\citep[SINGS,][]{ref:Kennicutt_2003} 
made a comprehensive mid-IR imaging and 
spectroscopic survey of 75 galaxies, many of them within $10\;$Mpc. The 
Local Volume Legacy Survey~\citep[LVL,][]{ref:Dale_2009} surveyed a total of 256 nearby galaxies, including 
all known galaxies inside a sub-volume bounded by $3.5\;$Mpc and an unbiased 
sample of S-Irr galaxies within a larger, and more representative, $11\;$Mpc 
sphere. The ongoing Spitzer Survey of Stellar Structure in Galaxies 
\citep[$S^4G$,][]{ref:Sheth_2008} is collecting data for $\sim2300$ 
galaxies within $40\;$Mpc using the warm \textit{Spitzer} ($3.6$ and $4.5 
\micron$) bands. 

For our paper, we selected 7 galaxies spanning a range of 
mass, morphology, distance, and star formation history.
Since this is a pilot study, the sample is deliberately eclectic rather than 
focused on a sample maximizing the star formation rate per galaxy. 
Ultimately we would like to examine all nearby galaxies rather than just a few. 
Table\,\ref{tab:galaxies} summarizes the properties of the targeted galaxies. 
The absolute magnitude $M_B$ and H$\alpha$ luminosity L(H$\alpha$) are from 
\citet{ref:Kennicutt_2008}, and L(H$\alpha$) is converted to star formation rate (SFR)
following Equation\,2 of \citet{ref:Kennicutt_1998b}. The foreground Galactic 
extinctions are from \citet{ref:Schlafly_2011}. The targeted galaxies have 
an integrated SFR of $\sim2\,\,$M$_\odot\,$\,year$^{-1}$.
For the assumed Salpeter IMF of \citet{ref:Kennicutt_1998b}, we can convert this to the massive ($M>8\,\,$M$_\odot$)
star formation rate of $\sim0.014$\,year$^{-1}$. The observed ccSN rate over that 20 years 
is $\sim0.15$\,year$^{-1}$ ($0.05 < R_{SN} < 0.35$\,year$^{-1}$, at 90\% confidence).
Since the ccSN rate should agree with the massive-star formation rate, 
this is a significant discrepancy for which we  have no obvious explanation,
and such mismatches are also found in other contexts~\citep[e.g.,][]{ref:Horiuchi_2011}. 

The nearest of our targeted galaxies, NGC\,6822 \citep[D\,$\simeq0.46$\,Mpc,][]{ref:Gieren_2006}, 
is a barred irregular galaxy \citep{ref:RC3_1991}. We included this small galaxy in our sample 
as an interesting nearby test case for examining large numbers of smaller, lower metallicity systems. 
M\,33~\citep[D\,$\simeq0.96$\,Mpc,][]{ref:Bonanos_2006} was previously 
studied by both \citet{ref:Thompson_2009} and \citet{ref:Khan_2010} to search 
for dusty stars that are much redder but less luminous than the stars 
we are searching for in this paper. 
NGC\,300 \citep[D\,$\simeq1.9$\,Mpc,][]{ref:Gieren_2005} and M\,81 \citep[D\,$\simeq3.6$\,Mpc,][]{ref:Gerke_2012} 
were also studied by \citet{ref:Khan_2010}.
NGC\,2403 \citep[D\,$\simeq3.1$\,Mpc,][]{ref:Saha_2006} contains two sources sometimes 
classified as SN impostors, SN\,1954J and SN\,2002kg~\citep[see the review by][]{ref:vanDyk_2005}, but 
any star associated with SN\,1954J must be relatively low mass ($\sim 20 \,$M$_\odot$ rather than $\gtrsim 50 \,$M$_\odot$)
and shows no strong evidence for mid-IR emission, while SN\,2002kg is simply a luminous variable star with little 
mass loss~\citep[see][]{ref:Kochanek_2012a}. Unlike the other large galaxies we studied, NGC\,247 
\citep[D\,$\simeq3.6$\,Mpc,][]{ref:Madore_2009} is highly inclined. 
NGC\,7793 \citep[D\,$\simeq4.1$\,Mpc,][]{ref:Tully_2009} is the most distant galaxy studied.

For M\,33, we used the six co-added epochs of IRAC data from~\citet{ref:McQuinn_2007} 
that were used by~\citet{ref:Thompson_2009} and~\citet{ref:Khan_2010}, and 
the MIPS data retrieved from the \textit{Spitzer Heritage Archive}. 
For NGC\,300 and NGC\,247, we used the data from the LVL survey \citep{ref:Dale_2009}. For NGC\,6822, NGC\,2403, M\,81, 
and NGC\,7793, we used the data from the SINGS survey~\citep{ref:Kennicutt_2003}. We utilize the 
full mosaics available for each galaxy. Table\,\ref{tab:photometry} shows the different pixel scales 
of the images retrieved from the \textit{Spitzer} archive for M\,33, and those provided by the SINGS 
and LVL surveys for the other six galaxies.
\subsection{Candidate Selection}
\label{sec:selection}
\begin{figure*}[p]
\begin{center}
\includegraphics[angle=0,width=150mm]{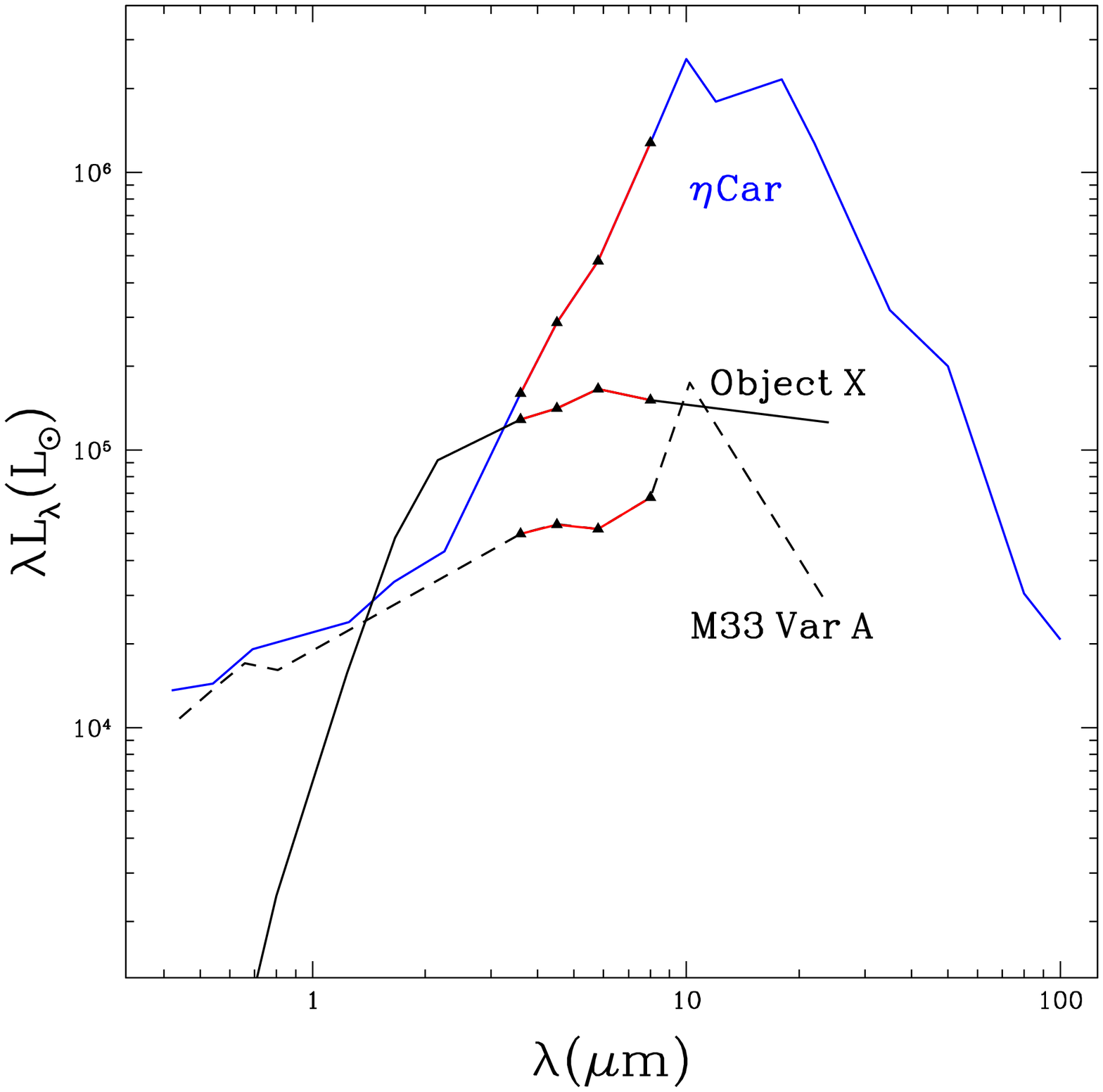}
\end{center}
\caption{The Spectral Energy Distributions (SED) of $\eta$\,Car now (blue solid line, \citealp{ref:Humphreys_1994}), 
Object\,X (black solid line, \citealp{ref:Khan_2011}), and M\,33\,Var\,A (black dashed line, \citealp{ref:Humphreys_2006}). 
The black triangles mark luminosity at the IRAC band centers. 
Although $\eta$\,Car and Object\,X have similar luminosities up to 3.6\,$\micron$, 
the SED of $\eta$\,Car is steeply rising in the IRAC bands ($a$\,$\simeq2.6$; Eqn.\,\ref{eqn:line_fit}) 
while Object\,X is almost flat ($a$\,$\simeq0.2$; Eqn.\,\ref{eqn:line_fit}). Object\,X, and M\,33\,Var\,A~\citep{ref:Hubble_1953,ref:Humphreys_2006} are 
both dust obscured stars with comparable bolometric luminosities~\citep{ref:Khan_2011}, 
but in the IRAC bands, Object\,X is much more luminous ($\sim1.5\times10^5 $L$_\odot$) 
than M\,33\,Var\,A ($\sim0.5\times10^5 $L$_\odot$).}
\label{fig:eta_sed}
\end{figure*}

The SED of a hot dust-obscured star will generally have two peaks --- a dust 
obscured optical peak, which could be absent altogether given enough absorption, and a mid-IR peak whose location in wavelength depends 
on the radius of the dust shell around the star. In the IRAC bands, the SED will be flat or rising towards 
longer wavelengths. For example, $\eta$\,Car has a steeply rising SED towards longer mid-IR 
wavelengths \citep[e.g.,][]{ref:Robinson_1973} and the luminosity of the star exceeds 
$10^5\,$\,L$_\odot$ in each IRAC band (see Figure\,\ref{fig:eta_sed}). At 
extra-Galactic distances, an $\eta$\,Car analog would appear as a bright, red point source 
in IRAC images with a relatively fainter optical counterpart due to the self-obscuration. 

We used the Daophot/Allstar PSF-fitting and photometry package 
\citep{ref:Stetson_1992} to identify point sources in all four IRAC bands 
and then performed photometry at the source location using both aperture 
and PSF photometry. 
We used the IRAF\footnote{IRAF is distributed by the 
National Optical Astronomy Observatory, which is operated by the Association of 
Universities for Research in Astronomy (AURA) under cooperative agreement with 
the National Science Foundation.} ApPhot/Phot tool for the aperture photometry. 
The aperture fluxes were transformed to Vega-calibrated magnitudes following the procedures 
described in the \textit{Spitzer} Data Analysis 
Cookbook\footnote{http://irsa.ipac.caltech.edu/data/SPITZER/docs/dataanalysistools/} 
and aperture corrections of 1.213, 1.234, 1.379, and 1.584 for the four IRAC bands. 
The choice of extraction aperture aperture ($R_{ap}$) as well as the inner ($R_{in}$) and outer 
($R_{out}$) radii of the local background annulus are reported in Table\,\ref{tab:photometry}. 
We estimate the local background using a $2\sigma$ outlier rejection procedure in 
order to exclude sources located in the local sky annulus, and correct for the 
excluded pixels assuming a Gaussian background distribution. Using a 
background annulus immediately next to the signal aperture minimizes the effects of 
background variation in the crowded fields of the galaxies. 
We used the Daophot/Allstar package for PSF photometry. 
The PSF photometry fluxes were transformed to Vega-calibrated magnitudes by 
applying zero point offsets determined from the difference between the calibrated aperture 
magnitudes and the initial PSF magnitude estimates of the bright stars in each galaxy.
\begin{table*}[t]
\begin{center}
\caption{Aperture Definitions}
\label{tab:photometry}
\begin{tabular}{cllrrr}
\\
\hline 
\hline
\multicolumn{1}{c}{Band} &
\multicolumn{2}{c}{Pixel Scale} &
\multicolumn{1}{c}{$R_{ap}$} &
\multicolumn{1}{c}{$R_{in}$} &
\multicolumn{1}{c}{$R_{out}$} 
\\
\multicolumn{1}{c}{($\mu$m)} &
\multicolumn{1}{c}{(Archive)} &
\multicolumn{1}{c}{(Survey)} &
\multicolumn{3}{c}{} 
\\
\hline
\hline
3.6-8.0 & $1\farcs2$ & $0\farcs75$ & $2\farcs4$ & $2\farcs4$ & $7\farcs2$ \\
24 & $2\farcs45$ & $1\farcs5$ & $3\farcs5$ & $6\farcs0$ & $8\farcs0$ \\
70 & $4\farcs0$ & $4\farcs5$ & $16\farcs0$ & $18\farcs0$ & $39\farcs0$ \\
160 & $8\farcs0$ & $9\farcs0$ & $16\farcs0$ & $64\farcs0$ & $128\farcs0$ 
\\
\hline
\hline
\end{tabular}
\end{center}
\end{table*}

For the 3.6 and 4.5$\micron$ bands, after verifying consistency with the aperture magnitudes, we only use the Vega-calibrated PSF 
magnitudes. For 5.8\,$\micron$, we switch to aperture magnitudes when Allstar 
fails to fit the PSF to a point source at the location identified by Daophot due to the decreasing resolution.
PSF photometry performs very poorly at 8.0\,$\micron$, leading 
to both inaccurate photometry and many false sources because Daophot frequently 
splits up extended regions of PAH emission into spurious point sources. Thus, at 
8.0\,$\micron$ we only use aperture photometry at positions determined for sources identified in the 
other three bands. We do not use this band for building our initial source list. 

We define our initial source list as all point sources that have $\lambda L_\lambda > 10^4 $\,L$_\odot$ 
in any one of the 3.6, 4.5, and 5.8\,$\micron$ bands, excluding regions near saturated 
stars and, in the case of M\,81, the high surface brightness core of the galaxy. 
We identify sources in each of these three bands, and cross-match the
catalogs using a 1\,pixel matching radius. We then adopt the position determined at the 
shortest wavelength (highest resolution) with a $>3\sigma$ detection, and we
use this position for the 8.0\,$\micron$ aperture photometry.
We fit the mid-IR SED of each object as a power law in wavelength 
\begin{equation}
\label{eqn:line_fit}
\log_{10}(\lambda L_\lambda) = a\times \log_{10}(\lambda) + b
\end{equation}
to determine the slope ($a$, $\lambda L_\lambda \propto \lambda^a$) and intercept ($b$). 
We can crudely relate the slope ($a$) to a dust temperature as 
\begin{equation}
\label{eqn:slope}
a = -4 + \frac {\log_{10} \left( \frac{e^{\frac{hc}{\lambda_1 k T}}-1}{e^{\frac{hc}{\lambda_4 k T}}-1} \right)} {\log_{10} \left( \frac{\lambda_4}{\lambda_1} \right) },
\end{equation}
where $\lambda_1$ and $\lambda_4$ are the shortest and longest band-centers 
assuming a blackbody spectrum and ignoring Planck factors. We 
define the total mid-IR luminosity ($L_{mIR}$) as the 
trapezoid rule integral of $L_\lambda$ across the band centers
\begin{equation}
\label{eqn:trapezoid}
L_{mIR} = \sum_{i=1}^{3} \frac{1}{2}\left( \lambda_{i+1} - \lambda_{i}\right) \left( L_{\lambda_{i}} + L_{\lambda_{i+1}}\right),
\end{equation}
where $\lambda_i=$ 3.6, 4.5, 5.8, and 8.0\,$\micron$.
We also calculate the fraction $f$ of $L_{mIR}$ that is emitted in the first three IRAC bands. We define $f$
as the ratio of the energy emitted between 3.6 and 5.8\,$\micron$ (first two terms of the integral), 
to $L_{mIR}$ (all three terms of the integral).
The approximate values of $L_{mIR}$, $a$, and $f$ for $\eta$\,Car are 
$10^{5.65}\,$\,L$_\odot$, $2.56$, and $0.32$, 
and those for Object\,X are $10^{5.17}\,$\,L$_\odot$, $0.22$, $0.57$. 

\begin{figure*}[p]
\begin{center}
\includegraphics[angle=0,width=150mm]{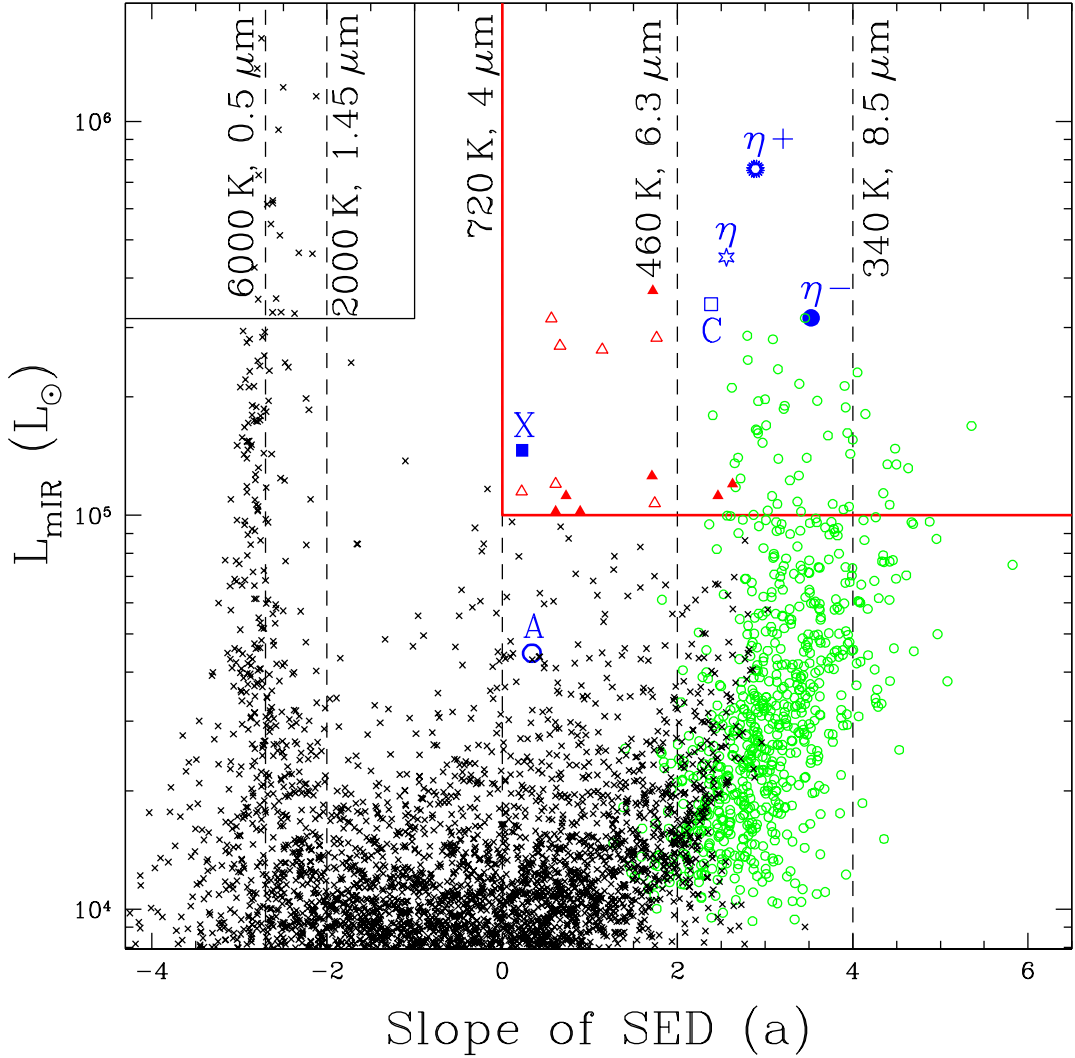}
\end{center}
\caption{\small{Integrated mid-IR luminosity $L_{mIR}$ as a function of the slope $a$ 
(Equation \ref{eqn:line_fit}) for bright 
sources in M\,81. The vertical dashed lines show the slopes of blackbodies 
with the indicated temperatures and peak wavelengths (Equation \ref{eqn:slope}). The 
top-right (thick red) box shows the candidate selection region 
($L_{mIR}>10^{5}\,$L$_\odot$ and $a>0$). The red triangles show the sources 
that also satisfy the third selection criteria, that at least 30\% of the 
integrated mid-IR luminosity is emitted between 3.6 and 5.8\,$\micron$ 
($f>0.3$). Of these, the open red triangles correspond to candidates that are 
known to be non-stellar in nature (see Sections \ref{sec:verify} and 
\ref{sec:other}), and the solid red triangles represent the surviving 
candidates. The green open circles show sources with $f<0.3$ and the black 
cross marks represent all the other sources. The narrow 
clump of points at $a\simeq-2.75$ correspond to normal stars with steeply 
falling mid-IR SEDs, while the wider clump of points to the right correspond 
to sources dominated by 8\,$\micron$ PAH emission. 
The top-left box shows the region $L_{mIR}>10^{5.5}\,$L$_\odot$ and $a<-1$ that was used to select 
normal stars in the M\,33 image (see Figure\,\ref{fig:m33_SEDs}).
The labeled blue points 
represent objects not in M\,81 that are shown for comparison: Object\,X (``X'', 
solid square), the compact cluster M\,33-8 (``C'', open square), 
M\,33\,Var\,A (``A'', large open circle), $\eta$\,Car 
(``$\eta$", open star), the Carina nebula excluding $\eta$\,Car 
itself \citep[``$\eta-$", solid circle;][]{ref:Smith_2007a}, and the 
Carina nebula including $\eta$\,Car (``$\eta$+", spiked open 
circle; see Section\,\ref{sec:clusters} and Figure\,\ref{fig:eta_complex}).}}
\label{fig:slope_lum}
\end{figure*}

\begin{figure*}[p]
\begin{center}
\includegraphics[angle=0,width=150mm]{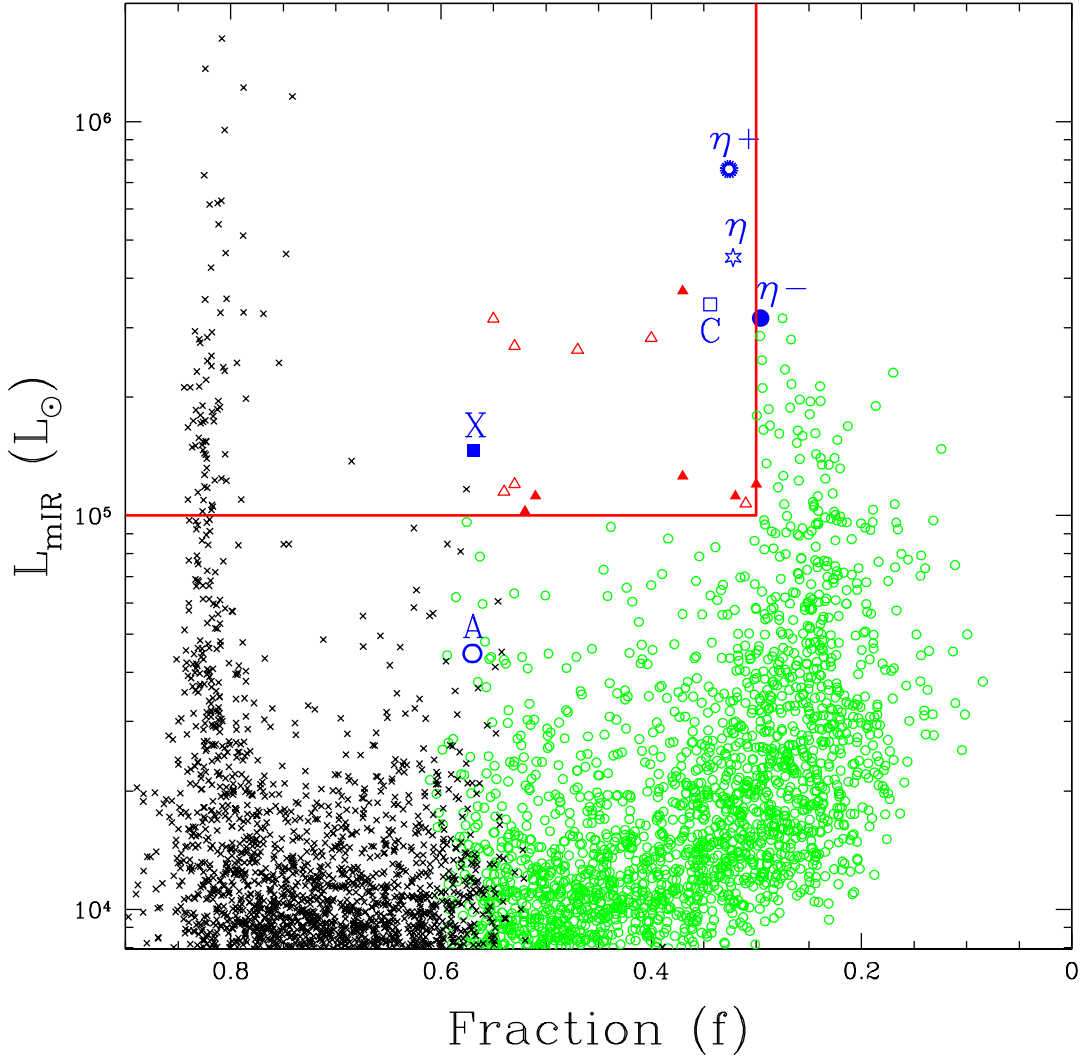}
\end{center}
\caption{Integrated mid-IR luminosity $L_{mIR}$ as a function of the fraction 
$f$ of $L_{mIR}$ that is emitted between 3.6 and 5.8\,$\micron$ for bright 
sources in M\,81. The box shows the candidate selection region 
($L_{mIR}>10^{5}\,$L$_\odot$ and $f>0.3$). The red triangles show the sources 
that also satisfy the third selection criteria that the mid-IR SED slope 
(Equations \ref{eqn:line_fit}) is either flat or rising 
($a>0$). Of these, the open red triangles correspond to candidates that are 
known to be non-stellar in nature (see Sections \ref{sec:verify} and 
\ref{sec:other}), and the solid red triangles represent the surviving 
candidates. The green open circles show sources with $a<0$ and the black cross 
marks represent all the other sources. The narrow clump of points at 
$f\simeq0.8$ correspond to normal stars with steeply falling (negative slope) 
mid-IR SEDs, while the wider clump of points at $f\simeq0.25$ correspond to 
sources dominated by 8\,$\micron$ PAH emission. The labeled blue points 
are same as in Figure\,\ref{fig:slope_lum}.}
\label{fig:frac_lum}
\end{figure*}

We defined candidates as sources with mid-IR luminosity $L_{mIR}>10^{5}\,$\,L$_\odot$, 
a mid-IR SED slope $a>0$, and $f>0.3$. Figures \ref{fig:slope_lum} 
and \ref{fig:frac_lum} show the distribution of point 
sources in M\,81 with $\lambda L_\lambda > 10^4 $\,L$_\odot$ in at least one of 
the 3.6, 4.5, and 5.8\,$\micron$ IRAC bands as a function of 
$L_{mIR}$, $a$, and $f$. The open red triangles in these figures correspond to 
candidates that are known to be non-stellar in nature (see Section\,\ref{sec:verify}), 
and the solid red triangles represent the surviving candidates. While a few 
hundred sources in M\,81 are bright enough in the mid-IR to be included in these figures, only 
a handful of these even remotely resemble $\eta$\,Car, and not a single 
one of them is as luminous and as red (cold) as $\eta$\,Car. The other targeted 
galaxies show similar distributions of sources. These distributions illustrate that 
our selection criteria for identifying potential $\eta$\,Car analogs 
are robust and allows for selecting objects that are significantly less luminous in the 
mid-IR and have much warmer circumstellar dust than $\eta$\,Car. Table\,\ref{tab:candidates}
reports the survey area and the number of candidates found for each galaxy.

\begin{figure*}[p]
\begin{center}
\includegraphics[angle=0,width=150mm]{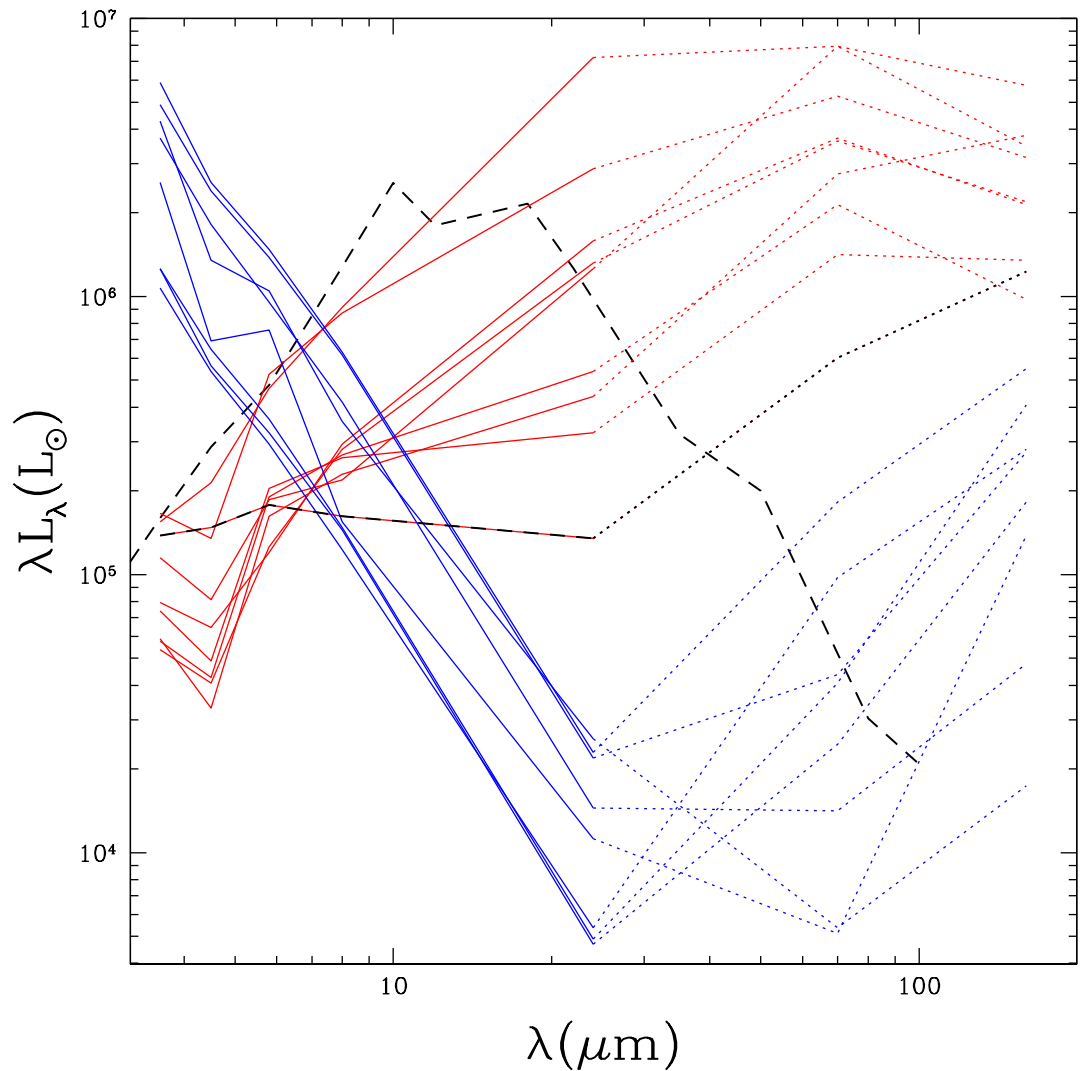}
\end{center}
\caption{Mid and far-IR SEDs of the candidates in M\,33 (red lines) compared to the SEDs of normal stars with $L_{mIR}>10^{5.5}$\,L$_\odot$, which steeply \textit{falling} SEDs (mid-IR slope $a<-1$, top left box of the Figure\,\ref{fig:slope_lum}). The dotted portions of the SEDs correspond to the MIPS 70 and 160\,$\micron$ flux upper limits. The SED of Object\,X is highlighted (red-black lighter dashed line) and $\eta$\,Car (black heavier dashed line) is shown for comparison.}
\label{fig:m33_SEDs}
\end{figure*}

We used aperture photometry to estimate the MIPS 24, 70, and 160\,$\micron$ band 
luminosities of the objects that meet our selection criteria. For point sources 
that do not have a flux that is $\gtrsim3\sigma$ above the local sky, we determine 
the $3\sigma$ detection limit for each aperture location using the local background 
estimate. Due to the poor spatial resolution of these bands, which forces us to 
choose increasingly large apertures at longer wavelengths (see Table\,\ref{tab:photometry}), 
these measurements have limited utility. Figure\,\ref{fig:m33_SEDs} shows the mid-IR 
SEDs of the candidates we identified in M\,33 along with normal stars in the M\,33 image
selected from top left region of Figure\,\ref{fig:m33_SEDs}. 
At 24\,$\micron$, the SEDs of the normal stars show the expected slope for the Rayleigh-Jeans tail 
of their SEDs, followed by an unphysical rise at 70 and 160\,$\micron$.
Essentially, due to the poor resolution, the apertures used for these two bands include 
many objects other than the intended target, and even normal stars appear to 
have rising far-IR SEDs. 
This means that we can generally use the 24\,$\micron$ fluxes while 
the 70 and 160\,$\micron$ measurements should be treated as upper limits 
regardless of their origin. Nevertheless, the MIPS bands are useful as a qualitative 
constraint on an object's physical nature (i.e. if it is a galaxy, QSO, cluster etc.). 
\subsection{Sources of Contamination}
\label{sec:verify}
While our selection criteria are designed to identify dust obscured individual 
stars, there are several classes of contaminating sources as well. 
QSOs have red mid-IR SEDs compared to stars \citep[e.g.,][]{ref:Stern_2005}, as do star 
forming galaxies with strong PAH emission at 8.0\,$\micron$ (e.g. the SED models 
in \citealp{ref:Assef_2010}). Sources in the 
galaxies such as dusty star clusters and \ion{H}{2} regions can also appear as candidates. 
We used the SIMBAD\footnote{http://simbad.u-strasbg.fr/} 
and VizieR\footnote{http://vizier.u-strasbg.fr/} services to search for previous 
classifications and near-IR counterparts from the 2MASS point source catalog~\citep{ref:Cutri_2003}.
We also noted other significant pieces of information, such as a candidate location coinciding 
with known radio and X-ray sources. For example,
many of the candidates in M\,33 are associated with known supernova remnants (SNRs).
We reject a source as ``non-stellar'' if it is a galaxy or QSO with a measured redshift or 
if archival images clearly show that it is a galaxy. 
The rejected sources that meet our selection criteria are described in Section\,\ref{sec:other}.

\begin{figure*}[p]
\begin{center}
\includegraphics[angle=0,width=150mm]{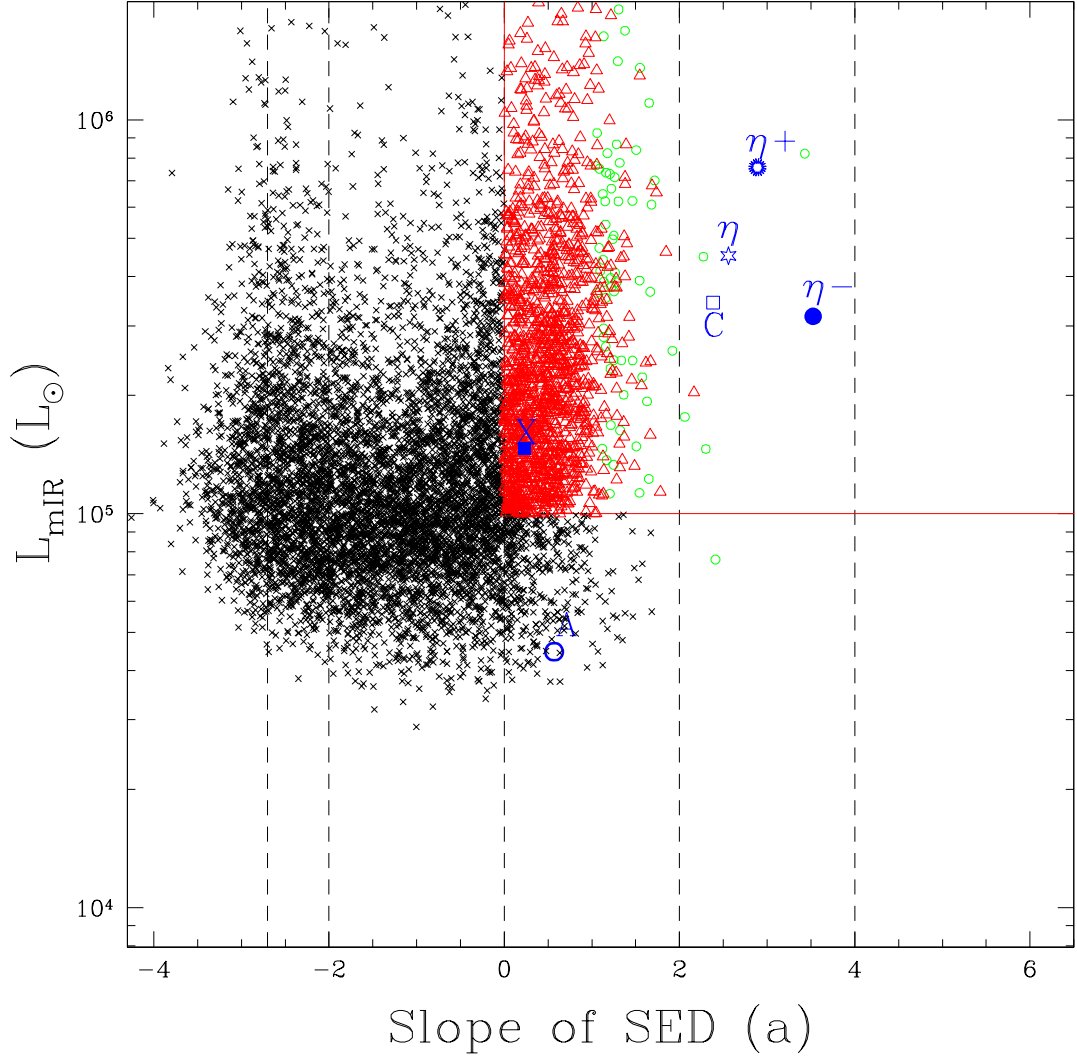}
\end{center}
\caption{Extragalactic contamination for M\,81. Here we show all sources from a 6\,deg$^2$ region of the SDWFS survey transformed to the distance of M\,81. The symbols, lines, and axis-limits are the same as in Figure\,\ref{fig:slope_lum}. In this SDWFS region, 449 ($\sim75$\,deg$^{-2}$) sources pass our selection criteria, indicating that we should expect $\sim13$ background sources meeting our selection criteria given our 0.17\,deg$^2$ survey region around M\,81. Note that very few of the contaminating background sources have properties comparable to $\eta$\,Car.}
\label{fig:bootes_slope_lum}
\end{figure*}

We estimate the expected number of extragalactic contamination for each galaxy 
using the SDWFS survey~\citep{ref:Ashby_2009} where the nature of the sources,
particularly AGNs, is also well understood from the AGES redshift survey~\citep{ref:Kochanek_2012b}. We transform the apparent 
magnitudes of all sources in a 6\,deg$^2$ region of SDWFS 
to luminosity using each target galaxy's distance modulus, determine 
how many of them would meet our selection criteria, and correct that 
count for our survey area around each galaxy. Table\,\ref{tab:bootes} reports 
the expected surface density of extragalactic contaminants and the number 
expected given the survey area around each galaxy. We expect a total of 
$\sim41$ extragalactic sources to pass our selection criteria across the targeted galaxies, 
as compared to 46 initial candidates. 
Figure\,\ref{fig:bootes_slope_lum}, which has the same format as Figure\,\ref{fig:slope_lum}, 
illustrates this for M\,81's distance. 
In the 6\,deg$^2$ SDWFS area, 449 ($\sim75$\,deg$^{-2}$) sources pass our selection criteria, indicating that we should 
expect $\sim13$ background sources meeting our selection criteria given our 0.17\,deg$^2$ 
survey region around M\,81, as compared to the 14 initial candidates selected.
Indeed, as we discuss in Section\,\ref{sec:discussion}, we can already identify 
11 of the 46 initial candidates as extragalactic. Statistically, this means that only $6\pm6$ are likely 
associated with the galaxies. Also note in Figure\,\ref{fig:bootes_slope_lum} that none of the 
contaminating background sources have properties directly comparable to $\eta$\,Car. 
The expected numbers of contaminating sources are generally consistent with the observed 
numbers with the exception of NGC\,247, which we investigated but appears to surely be a statistical fluke. 
The angular distribution of the candidates relative to the galaxies is also strongly suggestive 
of a dominant contribution for background sources.
\begin{table*}[t]
\begin{center}
\caption{Candidate Statistics}
\label{tab:bootes}
\begin{tabular}{lrrrrrrr}
\\
\hline 
\hline
\multicolumn{1}{c}{} &
\multicolumn{1}{c}{NGC} &
\multicolumn{1}{c}{M\,33} &
\multicolumn{1}{c}{NGC} &
\multicolumn{1}{c}{NGC} &
\multicolumn{1}{c}{M\,81} &
\multicolumn{1}{c}{NGC} &
\multicolumn{1}{c}{NGC} 
\\
\multicolumn{1}{c}{} &
\multicolumn{1}{c}{6822} &
\multicolumn{1}{c}{} &
\multicolumn{1}{c}{300} &
\multicolumn{1}{c}{2403} &
\multicolumn{1}{c}{} &
\multicolumn{1}{c}{247} &
\multicolumn{1}{c}{7793} 
\\
\hline 
\hline
Survey Area $A$ (deg$^2$) & 0.1 & 0.73 & 0.17 & 0.12 & 0.17 & 0.2 & 0.044 \\
Candidates & 0 & 9 & 1 & 5 & 14 & 3 & 14 \\
Expected Background $\Sigma$ (deg$^{-2}$) & 0.17 & 1.67 & 11 & 47 & 75 & 75 & 110 \\
Expected Contamination $A\Sigma$ & 0 & 1 & 2 & 6 & 13 & 15 & 5 \\
Rejected Candidates & 0 & 0 & 0 & 0 & 7 & 1 & 3 \\
Remaining Candidates & 0 & 9 & 1 & 5 & 7 & 2 & 11 \\
\hline
\hline
\end{tabular}
\end{center}
\end{table*}

Many of the candidate SEDs show a ``dip'' from 
3.6\,$\micron$ to 4.5\,$\micron$ before rising again at 5.8\,$\micron$ (see Figure\,\ref{fig:m33_SEDs}). This is 
a common feature of star cluster SEDs created by strong PAH emission at 3.6\,$\micron$~\citep{ref:Whelan_2011}. 
In total galaxy spectra, this is a weaker effect and the dominant PAH 
emission feature is at $8\micron$ and comes more from the diffuse ISM rather than individual stars or clusters.
The SEDs of $\eta$\,Car and Object\,X do not show this dip at 4.5\,$\micron$. 
We treat the presence of this dip as a qualitative indicator that the source 
may be a cluster or lie in a cluster. Deep \textit{Hubble Space Telescope} (\textit{HST}) images of these regions, 
where available, can help us distinguish single bright red stars from clusters of fainter stars 
that may be merged into a single bright source in the lower resolution \textit{Spitzer} images. 
Whether these clusters can potentially hide 
$\eta$\,Car analogs is discussed in Section\,\ref{sec:clusters}.

\begin{figure*}[p]
\begin{center}
\includegraphics[angle=0,width=150mm]{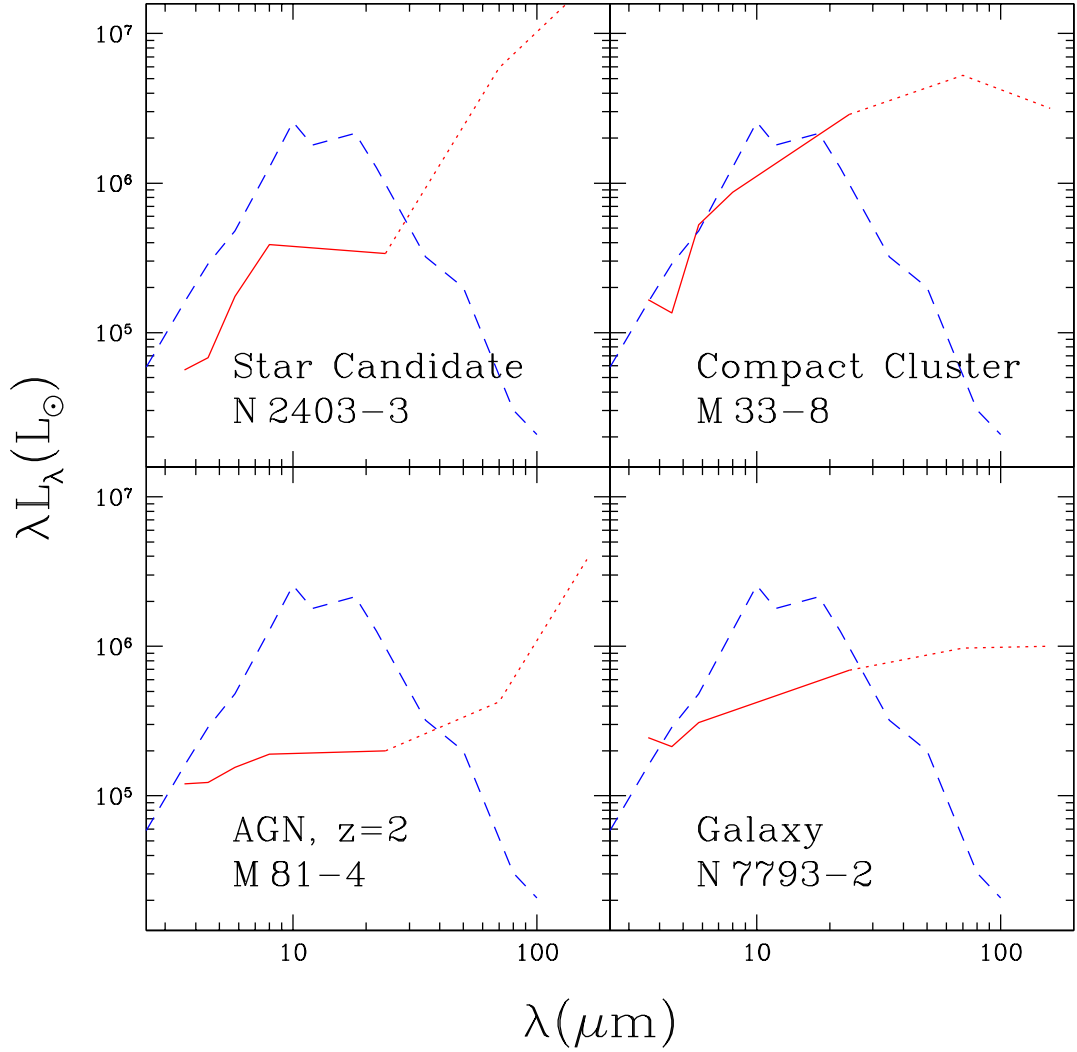}
\end{center}																	
\caption{SEDs of four different 
classes of objects that met our selection criteria: a candidate dusty star in NGC\,2403, 
a star-cluster in M\,33, a QSO behind M\,81, and a galaxy behind NGC\,7793. Figure\,\ref{fig:hst} 
shows IRAC and HST images of the compact cluster and the galaxy.}
\label{fig:example_4}
\end{figure*}

In Figure\,\ref{fig:example_4}, we present SEDs of four different 
types of objects that met our selection criteria --- a likely dusty star in NGC\,2403, 
a star-cluster in M\,33, a QSO behind M\,81, and a galaxy behind NGC\,7793. Although 
all four objects met our selection criteria, the detailed shapes of their SEDs are very 
different from each other. The likely stellar source, N\,2403-3, has a 
very steeply rising mid-IR SED that peaks between 8\,$\micron$ and 24\,$\micron$. 
While the compact cluster (M\,33-8) SED looks quite similar to that of $\eta$\,Car 
in the IRAC bands, it continues to rise steeply up to 24\,$\micron$, and 
the MIPS 70\,$\micron$ and 160\,$\micron$ upper limits show that it 
peaked between 24\,$\micron$ and 70\,$\micron$. The SED of the QSO (M\,81-4) remains relatively 
flat from 3.6\,$\micron$ to 24\,$\micron$. The MIPS 70\,$\micron$ and 
160\,$\micron$ upper limits for N\,7793-2 (a galaxy) are quite stringent, because the source 
is far from the center of the galaxy, and would rule out an $\eta$\,Car analog model.
For some cases, such as N\,7793-2, HST images clearly determine the nature of the source 
(Figure\,\ref{fig:hst}, bottom panel).
\subsection{Star Clusters}
\label{sec:clusters}
\begin{figure*}[p]
\begin{center}
\includegraphics[angle=0,width=150mm]{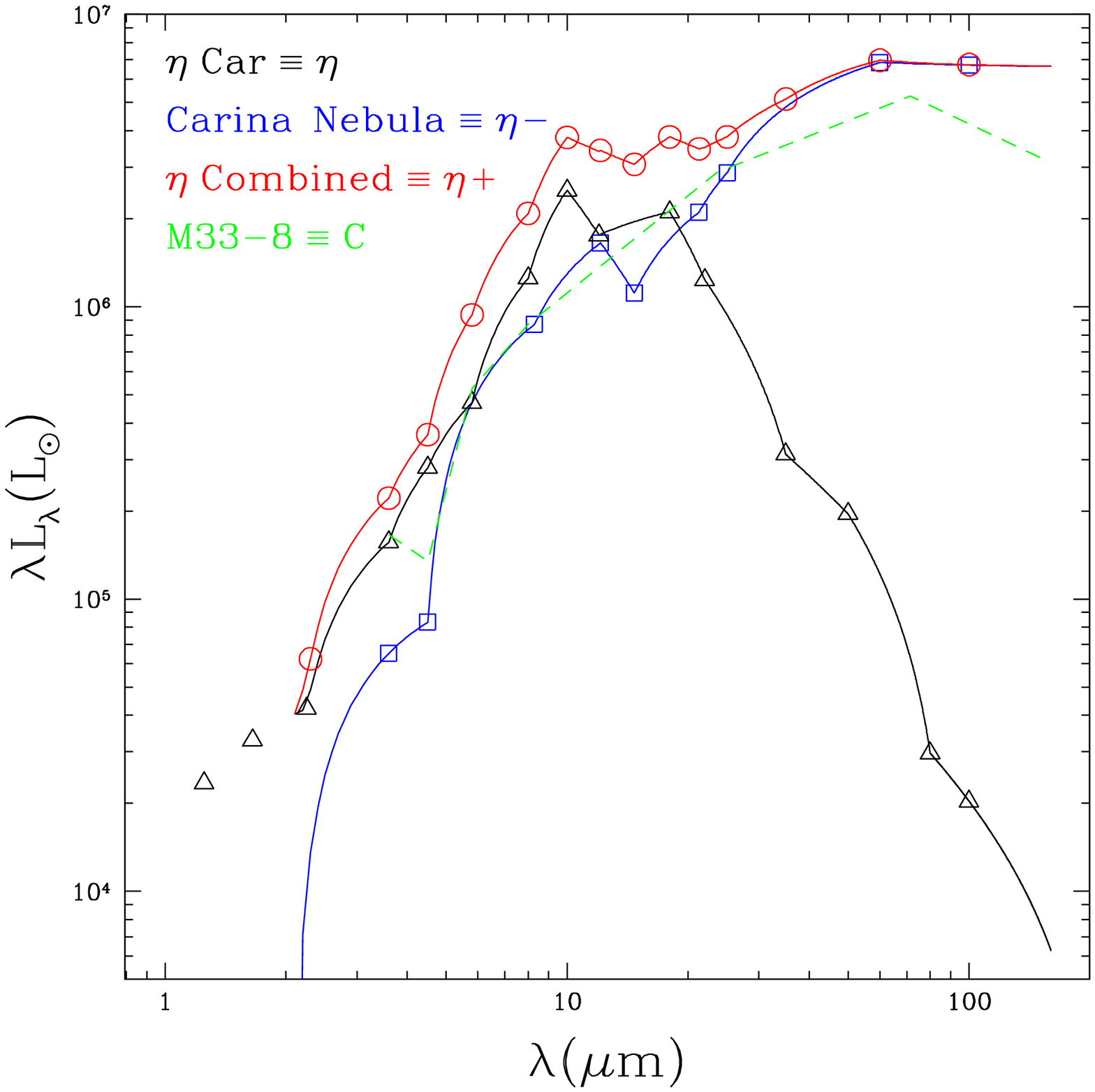}
\end{center}
\caption{The SEDs of $\eta$\,Car (``$\eta$'', black triangles, \citealp{ref:Humphreys_1994}), the Carina nebula excluding $\eta$\,Car itself (``$\eta$-'', blue squares, \citealp{ref:Smith_2007a}, and the entire dusty complex containing $\eta$\,Car and other massive stars including $\eta$\,Car (``$\eta$+'', red circles, Section\,\ref{sec:clusters}). The first two SEDs are spline interpolated and summed to produce the third. The SED of the compact cluster M\,33-8 (``C'', green dashed line, HST image in Figure\,\ref{fig:hst}) is shown for comparison. In Figures \ref{fig:slope_lum}, \ref{fig:frac_lum}, and \ref{fig:bootes_slope_lum} we label these $\eta$, $\eta-$, $\eta+$, and ``C'' respectively. }
\label{fig:eta_complex}
\end{figure*}

One concern with star clusters as a source of contamination 
is the possibility of ``hiding'' a luminous dusty 
star in a dusty star cluster. To explore this problem we estimated what the SED 
of the star cluster containing $\eta$\,Car would look like 
if it were located in one of the targeted galaxies. We combined the SED 
of $\eta$\,Car from \citet{ref:Humphreys_1994} with the SED of the 
Carina nebula from \citet{ref:Smith_2007a} to produce an SED of the entire 
complex (Figure\,\ref{fig:eta_complex}). The combined SED is clearly a 
multi-component SED, which is not typical of our candidates. Moreover, 
the Carina nebula is roughly $\sim 2.5\arcdeg$ in extent \citep{ref:Smith_2007a}, 
which at the distance of M\,33 becomes $\sim 20\farcs$ and would be easily resolved by IRAC.
Even at the distance of NGC\,7793 it would still subtend $\sim 5\farcs$
and be resolved. At all these distances it would be very easily 
resolved by HST or JWST~\citep{ref:Gardner_2006}.

In Figures \ref{fig:slope_lum} and \ref{fig:frac_lum}, we show the mid-IR luminosity $L_{mIR}$, 
SED slope $a$, and fraction $f$ of $\eta$\,Car (``$\eta$''), the Carina nebula excluding $\eta$\,Car (``$\eta-$''), 
and the entire complex including $\eta$\,Car (``$\eta+$''). It is apparent from these figures that 
even if the Carina nebula was not resolved: (1) we would select analogs of $\eta$\,Car and 
unresolved dusty stellar complexes hosting such analogs, (2) while it is close, we would not 
select a stellar complex that is identical to the Carina nebula excluding $\eta$\,Car, 
and (3) there are no sources with $L_{mIR}$, $a$, and $f$ comparable 
to $\eta$\,Car in M\,81. Indeed, this last point is true for each galaxy we studied.

There are, however, far more compact star clusters among the candidates such as M\,33-5, M\,33-8 
and M\,81-10 (see Section\,\ref{sec:candidates}) where HST images are required to recognize their spatial extent.
Even in these cases it is unlikely we would lose a candidate. First, it 
would require a ``conspiracy'' of a sort, namely that the SED of the hotter circumstellar dust
around the star (with characteristic $T\sim 400$\,K and $L_{bol}\sim few \times 10^6\,L_{\odot}$) 
seamlessly merges with the colder SED of the interstellar dust (with characteristic $T\sim100$\,K and 
$L_{bol}\sim 10^7\,L_{\odot}$) in the cluster. Typically we find that this leads to 
SEDs with ``bumps'' which we do not observe.

Possibly more constraining is the requirement that for a compact
cluster to hide an $\eta$\,Car analog it must still contain large 
amounts of interstellar gas and dust several million years after the 
cluster formed to allow for the time that even the most massive stars 
require to evolve away from the main sequence. However, a cluster sufficiently 
luminous to hide an $\eta$\,Car analog must host many luminous stars
with strong UV radiation fields and winds, which will likely clear the
cluster of gas and dust needed to produce strong
mid-IR emission. For example, 30\,Dor, 
which harbors stars possibly as massive as $300\,M_{\odot}$ and
is about $1.5\times 10^6\,$\,years old (e.g., \citealp{ref:Crowther_2010}), is
a weak source of $8\,\micron$ emission (see, e.g., Figure\,1 of \citealp{ref:Zhang_2012}).

\section{Inventory of Candidates}
\label{sec:candidates}
In this section we discuss the initial results of our search for $\eta$\,Car 
analogs. A total of 46 sources passed our basic mid-IR selection criteria 
($L_{mIR}>10^{5}\,\,$L$_\odot$, $a>0$, $f>0.3$). First we discuss the eleven candidates that can be 
rejected as known non-stellar sources. Their properties are reported in Table\,\ref{tab:others}. 
Then we list the remaining 35 candidates, including Object\,X, in Table\,\ref{tab:candidates}.
Table\,\ref{tab:extended} presents the near-IR photometry for the 9 sources 
with counterparts in the 2MASS point-source catalog~\citep{ref:Cutri_2003}. 
\subsection{Rejected Candidates}
\label{sec:other}
\begin{figure*}[p]
\begin{center}
\includegraphics[angle=0,width=150mm]{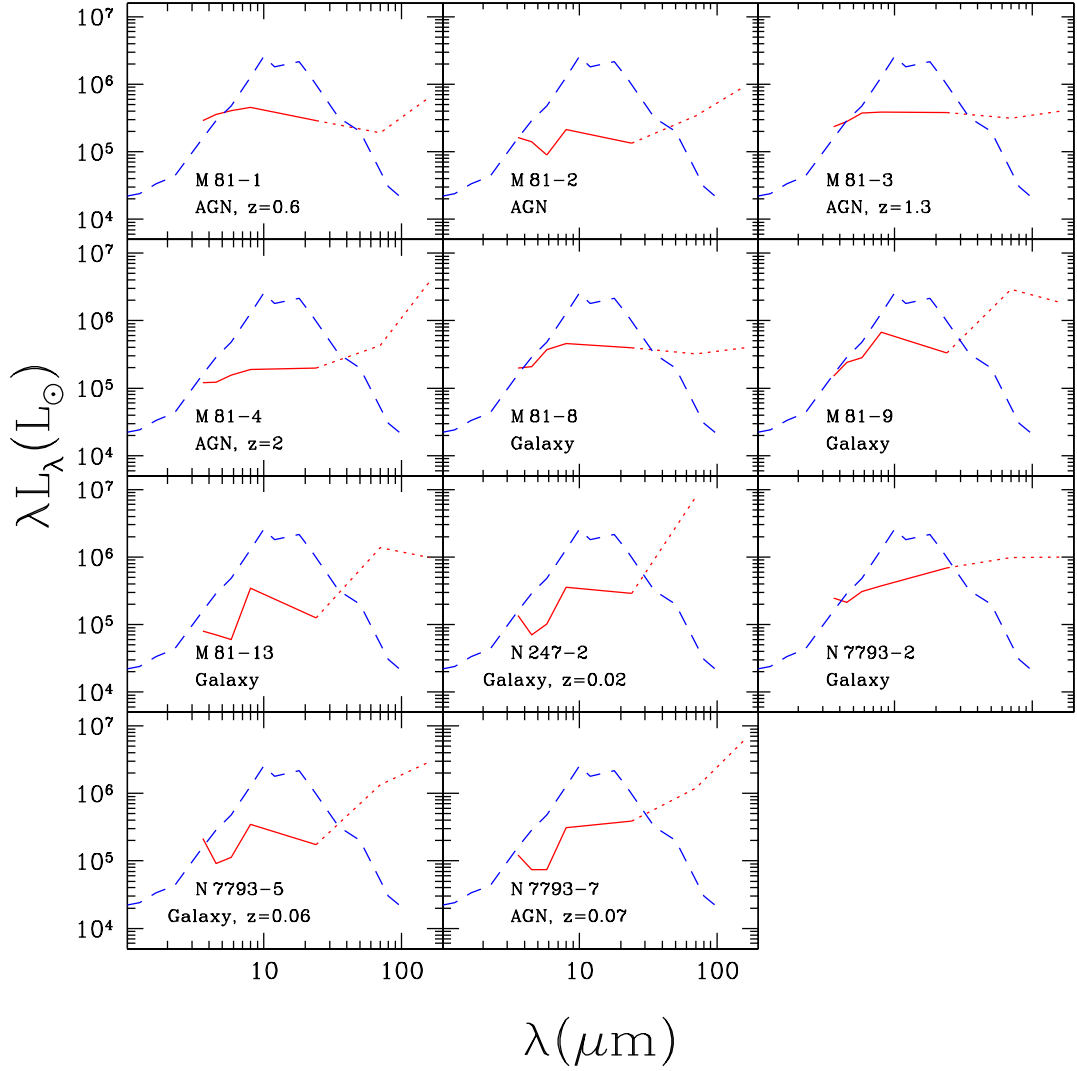}
\end{center}
\caption{SEDs of sources that met our selection criteria but were rejected due to association with non-stellar sources. The dotted portions of the SEDs correspond to the MIPS 70 and 160\,$\micron$ flux upper limits. The SED of $\eta$\,Car (dashed blue line) is shown for comparison.}
\label{fig:rejected}
\end{figure*}

\begin{table*}[!t]
\begin{center}
\begin{small}
\caption{Rejected Candidates}
\label{tab:others}
\begin{tabular}{rrrcccccccccc}
\\
\hline 
\hline
\multicolumn{1}{c}{ID} &
\multicolumn{1}{c}{RA} &
\multicolumn{1}{c}{Dec} &
\multicolumn{1}{c}{Slope} &
\multicolumn{1}{c}{\tiny{$\log_{10} { L_{mIR} \over L_\odot }$}} &
\multicolumn{1}{c}{$f$} &
\multicolumn{7}{c}{Spectral Energy Distribution [$\log_{10} (\lambda L_\lambda/$L$_\odot)$]} 
\\
\multicolumn{3}{c}{} &
\multicolumn{1}{c}{($a$)} &
\multicolumn{2}{c}{} &
\multicolumn{1}{c}{[3.6]} &
\multicolumn{1}{c}{[4.5]} &
\multicolumn{1}{c}{[5.8]} &
\multicolumn{1}{c}{[8.0]} &
\multicolumn{1}{c}{[24]} &
\multicolumn{1}{c}{[70]} &
\multicolumn{1}{c}{[160]} 
\\
\hline
\hline
M\,81-1  & 149.21474 & 69.12843  & 0.56  & 5.50  & 0.55  & 5.46  & 5.55  & 5.61 & 5.66  & 5.46 & $<$5.28 & $<$5.81 \\ 
M\,81-2  & 148.91331 & 69.29649  & 0.22  & 5.06  & 0.54  & 5.21  & 5.15  & 4.95 & 5.33  & 5.13 & $<$5.53 & $<$5.97 \\ 
M\,81-3  & 149.25626 & 68.91674  & 0.66  & 5.43  & 0.53  & 5.37  & 5.45  & 5.57 & 5.59  & 5.58 & $<$5.50 & $<$5.60 \\ 
M\,81-4  & 149.15222 & 69.00780  & 0.61  & 5.08  & 0.53  & 5.08  & 5.09  & 5.19 & 5.28  & 5.30 & $<$5.63 & $<$6.58 \\ 
M\,81-8  & 149.17365 & 68.80576  & 1.14  & 5.42  & 0.47  & 5.30  & 5.32  & 5.57 & 5.66  & 5.60 & $<$5.51 & $<$5.60 \\ 
M\,81-9  & 148.85034 & 69.24746  & 1.76  & 5.45  & 0.40  & 5.18  & 5.38  & 5.45 & 5.83  & 5.52 & $<$6.46 & $<$6.27 \\ 
M\,81-13  & 148.99657 & 69.26839  & 1.74  & 5.03  & 0.31  & 4.90  & 4.85  & 4.78 & 5.54  & 5.10 & $<$6.14 & $<$5.99 \\ 
N\,247-2  & 11.90806 & $-$20.51950  & 1.35  & 5.10  & 0.36  & 5.13  & 4.85  & 5.01 & 5.55  & 5.46 & $<$6.89 & \dots \\ 
N\,7793-2  & 359.47302 & $-$32.47820  & 0.62  & 5.37  & 0.51  & 5.39  & 5.33  & 5.49 & 5.57  & 5.84 & $<$5.99 & $<$6.00 \\ 
N\,7793-5  & 359.34467 & $-$32.62253  & 0.76  & 5.15  & 0.42  & 5.33  & 4.96  & 5.05 & 5.54  & 5.24 & $<$6.13 & $<$6.47 \\ 
N\,7793-7  & 359.46878 & $-$32.63801  & 1.17  & 5.04  & 0.37  & 5.09  & 4.87  & 4.87 & 5.49  & 5.59 & $<$6.08 & $<$6.80 
\\
\hline
\hline
\end{tabular}
\end{small}
\end{center}
\end{table*}
Of the 11 rejected candidates, six are AGNs or galaxies with a redshift 
measurement. Four have been photometrically classified as 
galaxies by the SDSS survey \citep{ref:Adelman_2009} and visual inspections of 
the SDSS images find diffuse extended sources consistent with this classification. 
One of these sources is also 0\farcs4 away from a radio and X-ray source and is 
likely a low redshift AGN. One candidate is a well-resolved galaxy in HST images. 
The SEDs of the rejected candidates are shown in Figure\,\ref{fig:rejected} 
and their luminosities, SED slope, and $f$ are listed in Table\,\ref{tab:others}.
In detail, we find that:
\begin{itemize}

\item{} \textit{M\,81-1} is an AGN. It is 0\farcs362 from a quasar at $z=0.605$~\citep{ref:Richards_2009}.
 
\item{} \textit{M\,81-2} is an AGN. It lies 0\farcs4 from a radio and X-ray source~\citep{ref:Flesch_2010} and is classified as a galaxy by 
SDSS~\citep{ref:Adelman_2009}. Visual inspection of the SDSS image also shows an extended source consistent with this classification.

\item{} \textit{M\,81-3} is an AGN. It is 0\farcs38 from a quasar at $z=1.29683$~\citep{ref:Schneider_2010}.

\item{} \textit{M\,81-4} is an AGN. It is 0\farcs65 from a quasar at $z=1.97519$~\citep{ref:Schneider_2010}.

\item{} \textit{M\,81-8}, \textit{M\,81-9}, and \textit{M\,81-13} are classified as galaxies by SDSS~\citep{ref:Adelman_2009} and visual inspections of the SDSS images finds extended sources consistent with these classifications.

\item{} \textit{N\,247-2} is a galaxy at $z=0.02089$~\citep{ref:Jones_2009}. 

\item{} \textit{N\,7793-2} is unambiguously a galaxy in HST images (Figure\,\ref{fig:hst}, bottom panel).

\item{} \textit{N\,7793-5} is a galaxy at $z=0.0614$~\citep{ref:Jones_2009}.

\item{} \textit{N\,7793-7} is an AGN. It is 1\farcs5 from a QSO at $z=0.071$~\citep{ref:Veron_2010}.
\end{itemize}

While the astrometric matches are sometimes imperfect, we are dealing with objects with such low surface densities that a mismatch is extraordinarily unlikely. Essentially, this search recapitulates aspects of the \citet{ref:Kozlowski_2010} search for quasars behind the Magellanic clouds as red mid-IR sources following the extragalactic mid-IR search criteria of \citet{ref:Stern_2005}. The red mid-IR colors created by the power-law SEDs of quasars mimic aspects of the red SEDs of dusty stars.
\subsection{Remaining Candidates}
\label{sec:promising}
\begin{table*}[!p]
\begin{center}
\begin{small}
\caption{Remaining Candidates}
\label{tab:candidates}
\begin{tabular}{rrrcccccccccc}
\\
\hline 
\hline
\multicolumn{1}{c}{ID} &
\multicolumn{1}{c}{RA} &
\multicolumn{1}{c}{Dec} &
\multicolumn{1}{c}{Slope} &
\multicolumn{1}{c}{\tiny{$\log_{10} { L_{mIR} \over L_\odot }$}} &
\multicolumn{1}{c}{$f$} &
\multicolumn{7}{c}{Spectral Energy Distribution [$\log_{10} (\lambda L_\lambda/\,$L$_\odot)$]} 
\\
\multicolumn{3}{c}{} &
\multicolumn{1}{c}{($a$)} &
\multicolumn{2}{c}{} &
\multicolumn{1}{c}{[3.6]} &
\multicolumn{1}{c}{[4.5]} &
\multicolumn{1}{c}{[5.8]} &
\multicolumn{1}{c}{[8.0]} &
\multicolumn{1}{c}{[24]} &
\multicolumn{1}{c}{[70]} &
\multicolumn{1}{c}{[160]} 
\\
\hline
\hline
M\,33-1\tablenotemark{a,b}  & 23.35015 & 30.42626  & 0.24  & 5.11  & 0.57  & 5.14  & 5.17  & 5.25 & 5.21  & 5.13 & $<$5.78 & $<$6.09 \\ 
M\,33-2\tablenotemark{a} & 23.39209 & 30.69071  & 1.06  & 5.10  & 0.46  & 5.06  & 4.91  & 5.27 & 5.34  & 6.10 & $<$6.90 & $<$6.76 \\ 
M\,33-3  & 23.43939 & 30.61357  & 1.98  & 5.10  & 0.39  & 4.87  & 4.69  & 5.31 & 5.42  & 5.51 & $<$6.15 & $<$6.13 \\ 
M\,33-4  & 23.55650 & 30.56175  & 2.17  & 5.01  & 0.36  & 4.77  & 4.52  & 5.21 & 5.36  & 5.64 & $<$6.44 & $<$6.58 \\ 
M\,33-5  & 23.31891 & 30.88054  & 2.32  & 5.57  & 0.36  & 5.19  & 5.33  & 5.67 & 5.96  & 6.86 & $<$6.90 & $<$6.54 \\ 
M\,33-6  & 23.36988 & 30.67363  & 1.79  & 5.05  & 0.36  & 4.90  & 4.81  & 5.08 & 5.47  & 6.20 & $<$6.57 & $<$6.33 \\ 
M\,33-7  & 23.39793 & 30.65805  & 2.33  & 5.08  & 0.36  & 4.76  & 4.63  & 5.28 & 5.43  & 5.73 & $<$6.33 & $<$5.99 \\ 
M\,33-8\tablenotemark{a}  & 23.50089 & 30.67987  & 2.40  & 5.56  & 0.35  & 5.22  & 5.13  & 5.72 & 5.94  & 6.46 & $<$6.72 & $<$6.50 \\ 
M\,33-9  & 23.37096 & 30.67276  & 2.34  & 5.02  & 0.32  & 4.73  & 4.61  & 5.10 & 5.45  & 6.12 & $<$6.56 & $<$6.34 \\ 
N\,300-1  & 13.71123 & $-$37.67159  & 0.96  & 5.27  & 0.50  & 5.17  & 5.27  & 5.37 & 5.50  & 5.28 & $<$6.23 & $<$6.30 \\ 
N\,2403-1\tablenotemark{a}  & 114.32964 & 65.59473  & 1.37  & 5.22  & 0.38  & 5.20  & 5.03  & 5.20 & 5.63  & 5.96 & $<$6.96 & $<$7.37 \\ 
N\,2403-2  & 114.20582 & 65.60922  & 2.00  & 5.11  & 0.36  & 4.93  & 4.68  & 5.25 & 5.49  & 5.31 & $<$7.03 & $<$7.57 \\ 
N\,2403-3  & 114.09702 & 65.61411  & 2.56  & 5.16  & 0.33  & 4.75  & 4.83  & 5.24 & 5.59  & 5.53 & $<$6.77 & $<$7.33 \\ 
N\,2403-4\tablenotemark{a}  & 114.22210 & 65.59257  & 2.20  & 5.17  & 0.32  & 4.98  & 4.71  & 5.24 & 5.61  & 5.24 & $<$6.71 & $<$7.45 \\ 
N\,2403-5  & 114.22632 & 65.59669  & 2.14  & 5.20  & 0.30  & 4.98  & 4.89  & 5.15 & 5.69  & 5.44 & $<$6.90 & $<$7.50 \\ 
M\,81-5  & 148.75421 & 69.12405  & 0.61  & 5.01  & 0.52  & 5.01  & 4.99  & 5.13 & 5.20  & 5.22 & $<$6.06 & $<$7.12 \\ 
M\,81-6  & 148.83128 & 68.95947  & 0.89  & 5.01  & 0.52  & 4.88  & 5.02  & 5.15 & 5.19  & 5.08 & $<$5.50 & $<$6.56 \\ 
M\,81-7  & 148.72035 & 69.14713  & 0.73  & 5.05  & 0.51  & 5.03  & 5.07  & 5.11 & 5.29  & 5.31 & $<$6.74 & $<$7.07 \\ 
M\,81-10  & 148.97075 & 68.98440  & 1.72  & 5.57  & 0.37  & 5.39  & 5.43  & 5.54 & 5.99  & 6.25 & $<$7.20 & $<$7.43 \\ 
M\,81-11  & 149.00545 & 68.98351  & 1.71  & 5.10  & 0.37  & 4.95  & 4.89  & 5.12 & 5.51  & 5.20 & $<$6.70 & $<$7.46 \\ 
M\,81-12  & 149.01479 & 68.98553  & 2.46  & 5.05  & 0.32  & 4.71  & 4.65  & 5.15 & 5.47  & 5.45 & $<$6.70 & $<$7.46 \\ 
M\,81-14  & 148.66461 & 69.08003  & 2.63  & 5.08  & 0.30  & 4.74  & 4.57  & 5.19 & 5.53  & 5.65 & $<$6.92 & $<$7.28 \\ 
N\,247-1  & 11.51449 & $-$20.72367  & 1.12  & 5.14  & 0.50  & 4.97  & 5.15  & 5.25 & 5.38  & 5.28 & \dots & $<$5.36 \\ 
N\,247-3\tablenotemark{a}  & 11.91676 & $-$20.90363  & 1.45  & 5.14  & 0.31  & 5.07  & 5.02  & 4.64 & 5.68  & 5.32 & $<$5.95 & $<$6.33 \\ 
N\,7793-1  & 359.39191 & $-$32.54715  & 0.31  & 5.13  & 0.55  & 5.20  & 5.28  & 4.95 & 5.40  & 5.21 & $<$5.45 & $<$6.47 \\ 
N\,7793-3  & 359.43268 & $-$32.60958  & 0.77  & 5.06  & 0.47  & 5.13  & 4.94  & 5.13 & 5.35  & 5.21 & $<$6.06 & $<$7.43 \\ 
N\,7793-4\tablenotemark{a}  & 359.38553 & $-$32.66690  & 0.91  & 5.34  & 0.46  & 5.34  & 5.32  & 5.31 & 5.67  & 5.53 & $<$5.98 & $<$6.36 \\ 
N\,7793-6\tablenotemark{a}  & 359.47568 & $-$32.60091  & 1.58  & 5.14  & 0.38  & 5.05  & 4.85  & 5.22 & 5.51  & 5.43 & $<$6.38 & $<$7.51 \\ 
N\,7793-8\tablenotemark{a}  & 359.42096 & $-$32.65040  & 1.65  & 5.21  & 0.33  & 5.20  & 4.83  & 5.19 & 5.67  & 5.24 & $<$6.09 & $<$6.68 \\ 
N\,7793-9  & 359.46133 & $-$32.58006  & 2.16  & 5.05  & 0.33  & 4.83  & 4.65  & 5.14 & 5.47  & 5.22 & $<$6.94 & $<$7.46 \\ 
N\,7793-10  & 359.41776 & $-$32.61156  & 2.23  & 5.19  & 0.32  & 4.99  & 4.68  & 5.30 & 5.61  & 5.54 & $<$6.67 & $<$7.43 \\ 
N\,7793-11  & 359.41071 & $-$32.60335  & 2.48  & 5.19  & 0.31  & 4.89  & 4.72  & 5.30 & 5.62  & 5.79 & $<$6.96 & $<$7.50 \\ 
N\,7793-12  & 359.48233 & $-$32.60726  & 2.34  & 5.02  & 0.31  & 4.73  & 4.65  & 5.06 & 5.47  & 5.54 & $<$6.46 & $<$7.65 \\ 
N\,7793-13  & 359.42245 & $-$32.59289  & 3.08  & 5.23  & 0.31  & 4.72  & 4.64  & 5.41 & 5.64  & 5.77 & $<$6.93 & $<$7.60 \\ 
N\,7793-14  & 359.47440 & $-$32.57985  & 2.73  & 5.16  & 0.30  & 4.78  & 4.61  & 5.29 & 5.59  & 5.48 & $<$6.59 & $<$7.32 
\\
\hline
\hline
\end{tabular}
\end{small}
\end{center}
\tablenotetext{a}{Identified as point sources in the 2MASS catalog~\citep{ref:Cutri_2003}. See Table\,\ref{tab:extended} for near-IR magnitudes.}
\tablenotetext{b}{M\,33-1 is Object\,X from \citet{ref:Khan_2011}.}
\end{table*}
\begin{table*}[!t]
\begin{center}
\begin{small}
\caption{Optical/NIR Luminosities of Stellar Candidates with 2MASS Counterparts}
\label{tab:extended}
\begin{tabular}{rrrrrrrrrrr}
\\
\hline 
\hline
\multicolumn{1}{c}{ID} &
\multicolumn{1}{c}{RA} &
\multicolumn{1}{c}{Dec} &
\multicolumn{1}{c}{$U$} &
\multicolumn{1}{c}{$B$} &
\multicolumn{1}{c}{$V$} &
\multicolumn{1}{c}{$R$} &
\multicolumn{1}{c}{$I$} &
\multicolumn{1}{c}{$J$} &
\multicolumn{1}{c}{$H$} &
\multicolumn{1}{c}{$K$} 
\\
\hline
\hline
M\,33-1 & 23.35015 & 30.42626 & $\gtrsim24.1$ & $\gtrsim24.2$ & 23.15 & 21.61 & 19.99 & 17.07 & 15.04 & 13.6 \\
M\,33-2 & 23.39209 & 30.69071 & 20.61   & 21.84   & 21.14 & 20.13 & 20.48 & 15.96 & 14.68 & 14.02 \\
M\,33-8 & 23.39026 & 30.69038 & 18.89   & 19.85   & 18.81 & 17.94 & 17.68 & 16.15 & 15.49 & 14.25 \\
N\,2403-1 & 114.32964 & 65.59473 & \dots   & \dots   & \dots & \dots & \dots & 14.89 & 14.42 & 14.21 \\ 
N\,2403-4 & 114.22210 & 65.59257 & \dots   & \dots   & \dots & \dots & \dots & 17.21 & 16.05 & 14.45 \\ 
N\,247-3 & 11.91676 & $-$20.90363 & \dots  & \dots   & \dots & \dots & \dots & 15.73 & 14.79 & 14.58 \\ 
N\,7793-4 & 359.38553 & $-$32.66690 & \dots  & \dots   & \dots & \dots & \dots & 16.30 & 15.59 & 15.17 \\ 
N\,7793-6 & 359.47568 & $-$32.60091 & \dots  & \dots   & \dots & \dots & \dots & 16.45 & 15.98 & 15.58 \\ 
N\,7793-8 & 359.42096 & $-$32.65040 & \dots  & \dots   & \dots & \dots & \dots & 16.07 & 15.56 & 14.14 
\\
\hline
\hline
\end{tabular}
\end{small}
\end{center}
\end{table*}
The mid-IR luminosities, slopes, and the fractions $f$ of the 35 remaining candidates, including Object\,X, 
are presented in Table\,\ref{tab:candidates}. Of these, 9 also have 2MASS photometry and 
their near-IR magnitudes are presented in Table\,\ref{tab:extended}
along with the $UBVRI$ magnitudes of the closest optical counterparts 
for the three candidates in M\,33 for which there is a publicly available 
catalog~\citep{ref:Massey_2007}.
We select Object\,X~\citep{ref:Khan_2011} as a candidate (M\,33-1) but not 
M\,33\,Var\,A~\citep{ref:Hubble_1953,ref:Humphreys_2006}.
Although both are dust obscured stars with comparable bolometric 
luminosities and exist in our initial source list, 
it is apparent from Figure\,\ref{fig:eta_sed} that 
in the IRAC bands Object\,X is much more luminous 
($\sim1.5\times10^5 \,$L$_\odot$) than M\,33\,Var\,A ($\sim0.5\times10^5 \,$L$_\odot$). 
In fact M\,33\,Var\,A recently emerged from a $\sim50$\,year dust obscured phase 
and is fading in the IRAC bands~\citep{ref:Humphreys_2006}, and we would probably 
have selected it in observations at an earlier epoch. 

\begin{figure*}[p]
\begin{center}
\includegraphics[angle=0,width=175mm]{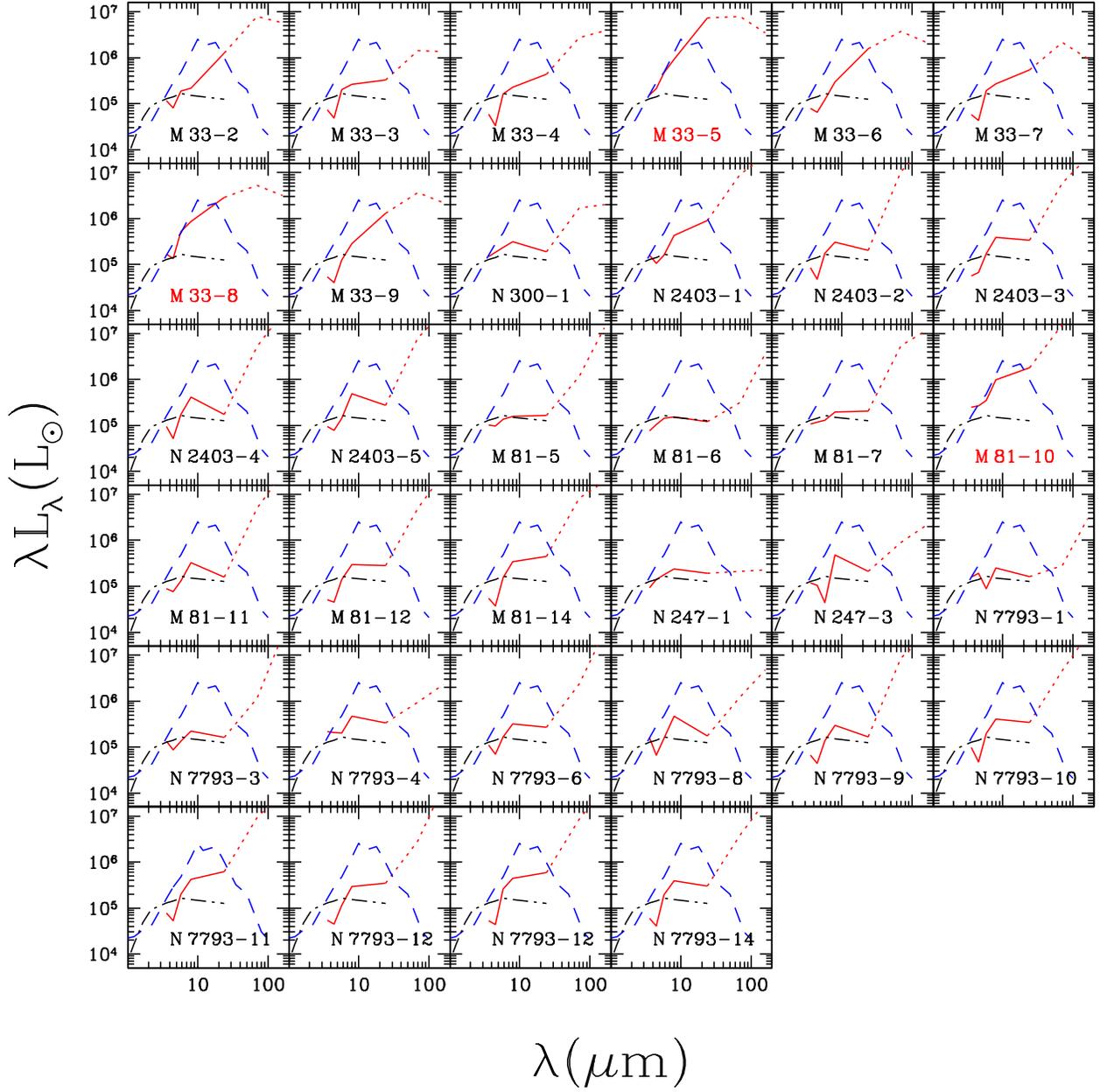}
\end{center}
\caption{SEDs of sources that met our selection criteria and were \textit{not} rejected due to association with non-stellar sources. The dotted portions of the SEDs correspond to the MIPS 70 and 160\,$\micron$ flux upper limits. The SEDs of $\eta$\,Car (dashed blue line) and Object\,X (dot-dashed black line) are shown for comparison.}
\label{fig:selected}
\end{figure*}

\begin{figure*}[p]
\begin{center}
\includegraphics[angle=0,width=105mm]{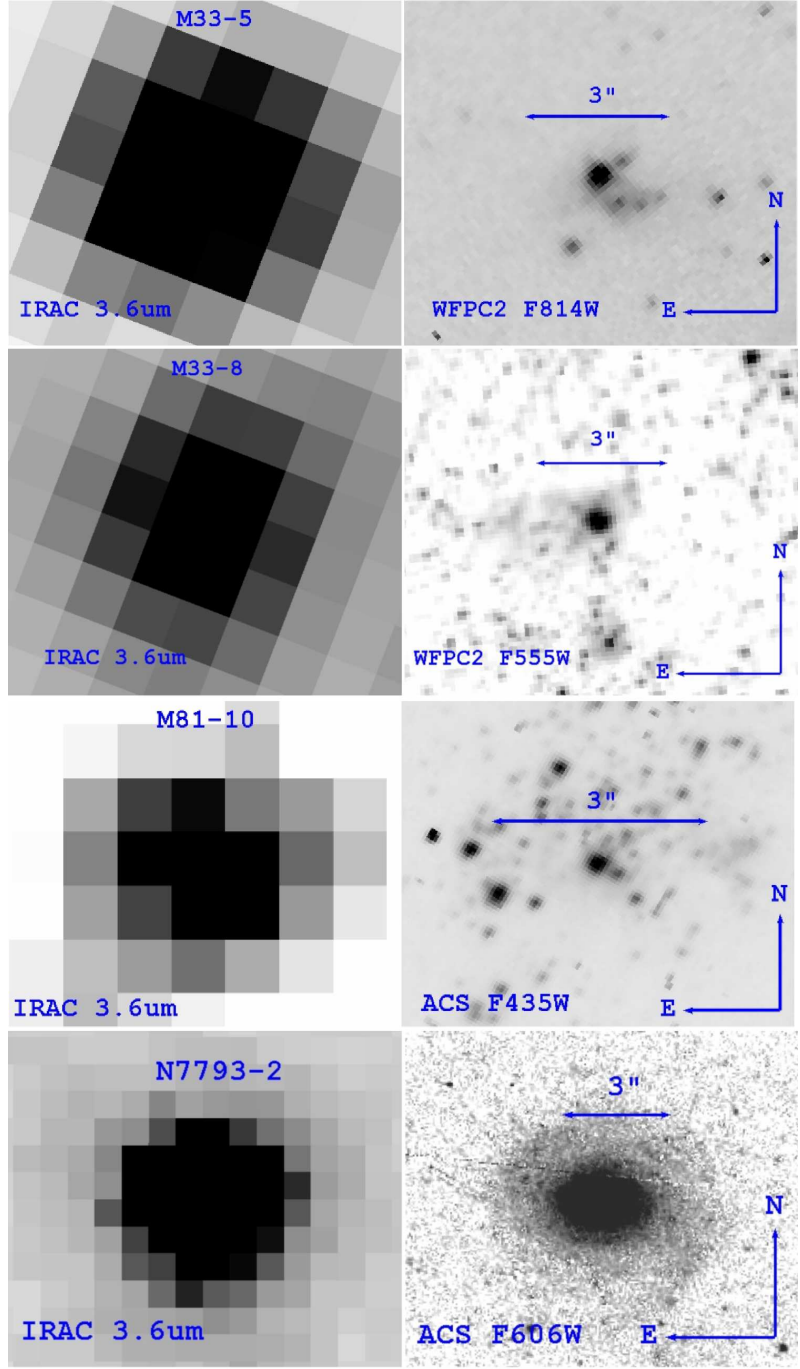}
\end{center}
\caption{IRAC and HST images of the compact stellar clusters M\,33-5, M\,33-8, and M\,81-10, and the background galaxy N\,7793-2. The clusters are resolved in the HST images with FWHM of $0\farcs87\simeq4.1$\,pc (M\,33-5), $0\farcs77\simeq3.6$\,pc (M\,33-8) and $0\farcs34\simeq6.1$\,pc (M\,33-8). They are very luminous (few$\times10^7\,$L$_\odot$) and their SED shapes are very similar to $\eta$\,Car (Figure\,\ref{fig:selected}).}
\label{fig:hst}
\end{figure*}

It is apparent from the SEDs of the candidates (Figures\,\ref{fig:selected}) 
that few have mid-IR luminosities comparable to $\eta$\,Car. Moreover, 
the SEDs of many sources, such as N\,2403-2, N\,247-3 and N\,7793-10, appear so 
dissimilar from the SEDs of $\eta$\,Car and Object\,X that they seem unlikely 
to be stellar sources. The MIPS band luminosity limits are also useful here. 
For example, the relatively low MIPS 70 and 160\,$\micron$ luminosity limits indicate 
that N\,247-1 must have a very flat or falling SED at these redder wavelengths and is probably an AGN, 
which is also suggested by its distance from the galaxy.

On the other hand, the SEDs of M\,33-5, M\,33-8 and M\,81-10 are similar to $\eta$\,Car
while N\,300-1, M\,81-5, M\,81-6 and M\,81-7 are similar to Object\,X (M\,33-1). 
HST images show that M\,33-5, M\,33-8, and M\,81-10 are compact star clusters 
(Figure\,\ref{fig:hst}, top three panels). M\,81-5 is 0\farcs56 from a variable X-ray source 
with maximum luminosity of $2\times10^{38}$\,ergs\,s$^{-1}$, which is consistent with the source 
being an X-ray binary~\citep{ref:Remillard_2006}. M\,81-7 has been classified as a galaxy 
by the SDSS survey but with a photometric redshift $z=0.00049$ comparable to that of M\,81 
\citep{ref:Adelman_2009}. HST images show a region of enhanced star formation consisting of at 
least two components. 
Ancillary data also shows that N\,7793-3 is a High-Mass 
X-ray Binary~\citep{ref:Mineo_2012} with maximum X-ray luminosity 
of $3.9\times10^{37}$\,ergs\,s$^{-1}$~\citep{ref:Liu_2011}. Although 
we do not currently have an explanation, 5 of the candidates in M\,33 
(M\,33-3 through M\,33-7) are within $\lesssim2\farcs0$ of radio-selected 
supernova remnants \citep{ref:Gordon_1999}. 
\section{Rate Limits}
\label{sec:discussion}
One advantage of searching for eruptions in the dust obscured phase is that the
process is relatively easy to simulate. We eject $M_e$ of material
from a star of luminosity $L_*$ and temperature $T_*$ at velocity $v_e$ over
time period $t_e$ and assume it forms dust with total (absorption
plus scattering) visual opacity $\kappa_V$ once it is sufficiently distant from the star. We 
can then use DUSTY~\citep{ref:Elitzur_2001} to simulate the evolution of the mid-IR 
luminosities and determine the time $t_d$ during which the source would
satisfy our selection criteria. Here we use $\kappa_V=84$~cm$^2~$g$^{-1}$,
roughly appropriate for silicate dust, but this is important only
to the extent that the ejecta mass can be rescaled as $M_e \propto \kappa_V^{-1}$.
The key variable for estimating rates is the expansion velocity $v_e$,
because the detection period scales as $t_d \propto v_e^{-1}$. The
velocities cited for the supernova impostors \citep[e.g.,][]{ref:Smith_2011}
and the velocity associated with the long axis of $\eta$ Car are
high, $v_e \gtrsim 500$~km\,s$^{-1}$. These velocities are very different
from those observed for the older, massive shells in the Galaxy
or the shorter axis of $\eta$ Car, where $v_e \lesssim 100$~km\,s$^{-1}$ (see
the discussion of this difference in \citealp{ref:Kochanek_2011b}). Here
we scale the results to $v_e=100$~km\,s$^{-1}$ since, for example, it 
results in our detecting systems with parameters similar to $\eta$
Car at its present age, as observed, and agrees with the expansion
velocities of the other massive Galactic shells around luminous stars. 
 
Detection of a shell at late times ($t_d \gg t_e$) is limited by its 
optical depth and temperature.
The shell has total visual optical depth greater than $\tau_V$ for 
\begin{equation}
 t(\tau_V) = \left( { M_e \kappa_V \over 4 \pi v_e^2 \tau_V}\right)^{1/2}
  \simeq 400 \tau_V^{-1/2} \left( { M_e \over M_\odot} 
   { \kappa_V \over 100~\hbox{cm}^2/\hbox{g} } \right)^{1/2} 
   \left( { 100~\hbox{km\,s$^{-1}$} \over v_e }\right)~\hbox{years},
\end{equation}
and once $\tau_V <1$ it begins to rapidly fade in the mid-IR. Ignoring
Planck factors, the spectral energy ($\lambda L_\lambda$) peaks at
\begin{equation}
 \lambda = { h c \over 4 k T_d } 
   \simeq 2 \left( {L_* \over 10^6 L_\odot}\right)^{1/4}
 \left( { t \over \hbox{year}} { v_e \over 100~\hbox{km\,s$^{-1}$} }\right)^{1/2}~\mu\hbox{m},
\end{equation}
so the emission peak shifts out of the IRAC bands after several decades,
and our survey is primarily limited by the shift of the emission
to longer wavelengths rather than the declining optical depth. 
It is better to search for these sources at 24\,$\micron$ as has been done in 
the galaxy~\citep{ref:Wachter_2010,ref:Gvaramadze_2010} but that would 
require the resolution of JWST~\citep{ref:Gardner_2006}.
A reasonable power-law fit to the results ($-1 \leq \log{M_e/M_\odot} < 1$,
$5.5 < \log(L_*/L_\odot) < 6.5$) of the DUSTY models is that the detection period is

\begin{equation}
 t_d \simeq t_e + 66 
   \left( { 100~\hbox{km$s^{-1}$} \over v_e }\right)
   \left( { L_* \over 10^6 L_\odot }\right)^{0.82}
   \left( { M_e \over M_\odot }\right)^{0.043}~\hbox{years}.
\end{equation}
For $M_e\simeq 10$\,M$_\odot$ and L$_* \simeq 10^{6.5}$\,L$_\odot$ like 
$\eta$~Car, $t_d \simeq t_e + 190 (100~\hbox{km$s^{-1}$}/v_w)$~years
where $t_e$ may also be $50$~years or more (see the discussion
in \citealp{ref:Kochanek_2012a}). For present purposes, we 
adopt $t_d=200$~years as the period over which our selection
criteria would identify an analogue of $\eta$ Car, consistent
with the fact that our selection criteria do identify $\eta$
Car.
 
We can normalize the rate of eruptions to the ccSN rate as 
\begin{equation}
  R_{erupt} \simeq 0.1 \left( 40 M_\odot \over M_{erupt} \right)^{1.35}
  N_{erupt} R_{SN}
     = f_\eta R_{SN},
\end{equation}
where $R_{SN}$ is the supernova rate and all stars more massive than $M_{erupt}$
undergo $N_{erupt}$ eruptions. Following the rate arguments in 
\cite{ref:Kochanek_2011b}, we can estimate the number of eruptions
per massive star needed to explain the massive Galactic shells. 
If there are $N_{shell} \simeq 10$ massive Galactic shells 
associated with massive stars ($M>M_{erupt}$), then
\begin{equation}
  N_{erupt} \simeq 2 \left( { N_{shell} \over 10 }\right)
     \left( { \tau_V \over 0.01 } \right)^{1/2}
     \left( \hbox{century}^{-1} \over R_{SN,MW} \right)
     \left( { M_{erupt} \over 40 M_\odot } \right)^{1.35}
     \left( { 10 M_\odot \over M_e } \right)^{1/2}
     \left( { v_e \over 100~\hbox{km\,s$^{-1}$} } \right),
\end{equation}
where $\tau_V = 0.01$ is the minimum optical depth needed to detect a shell 
surrounding the star and $R_{SN,MW}\sim 1$/century is the Galaxy's 
supernova rate. Since the Galactic shells are identified as shells
primarily at $24\mu$m, they are easier to find at low optical depths
and temperatures than in our extragalactic survey.  
Thus, the massive Galactic shells imply an eruption
rate relative to the supernova rate of $f_\eta \gtrsim 0.2 $ 
since it is unclear whether
we possess a complete inventory. Note that with this normalization
the rate estimate does not depend on the mass scale $M_{erupt}$.
\section{Conclusions}
\label{sec:conclusions}
This work empirically demonstrates that true analogs of $\eta$\,Car --- 
massive stars that have undergone eruptive mass ejection in the 
recent past (centuries) --- are rare. 
Based on the discussion in Section\,\ref{sec:discussion}, our survey can detect close
analogs of $\eta$ Car for roughly $t_d \simeq 200$~years,
consistent with $\eta$\,Car meeting our selection criteria. 
The statistics of our present sample gives us a maximum of
$N_{cand}=6 \pm 6$ candidate systems after correcting for
the estimated extragalactic contamination. 
Aside from the three very compact, luminous 
star clusters, the candidates generally do not have SEDs that closely resemble the SED of $\eta$\,Car. 
Although we keep those three compact clusters in our candidate list for now, 
it is highly unlikely that they could hide luminous 
dusty stars similar to $\eta$\,Car. We anticipate that further 
analysis and follow-up observations, using HST astrometry and photometry, ground
based spectroscopy and Herschel 70\,$\micron$ photometry, will show that 
most, if not all, of the remaining candidates are either non-stellar or are not truly analogous 
to $\eta$\,Car. Of the true stellar systems, they are clearly
going to be a mixture of ``eruptions'' such as $\eta$\,Car and sources with longer lived, relatively steady
dusty winds such as Object\,X. 

We will carry out the detailed consideration of the candidates in Paper\,II,
but suppose we scale conclusions about the rates to $N_{cand}=3$,
which would also correspond to the 95\% confidence upper limit
we would use for estimating rate limits if we were to eliminate
all the remaining candidates. This implies that we are
probing eruption rates of order\begin{equation}
R_{erupt} = \frac{N_{cand}}{t_d} = 0.015 \left( \frac{N_{cand}}{3} \right) \left( \frac{200~\hbox{year}}{t_d} \right)~\hbox{year}^{-1} \end{equation}
for this sample of galaxies (roughly 1 per 200 years per galaxy), and fractional rates compared to the ccSN rate of order \begin{equation}
f_\eta = 0.15 \left( \frac{N_{cand}}{3} \right) \left( \frac{200~\hbox{year}}{t_d} \right) \left( \frac{0.1~\hbox{year}^{-1}}{R_{SN}} \right) \end{equation}
that are in the appropriate regime. In fact, it seems
likely that we should not find $N_{cand}=0$ at the end
of Paper\,II, and in some senses we already have that
$N_{cand}\gtrsim1$ since our present sample contains 
Object\,X. 

Alternatively, we could estimate the expected number of candidates 
from ccSN rate and the statistics of Galactic shells as discussed in Section\,\ref{sec:discussion}.
For the galaxies in our pilot study, we have two rather inconsistent
estimates of the ccSN rate. Empirically, there were three ccSN over
the last 20 years, which implies a rate of $R_{SN}=0.15$\,year$^{-1}$ 
($0.05 < R_{SN} < 0.35$\,year$^{-1}$, at 90\% confidence). On the other hand, 
the integrated star formation rate of the targeted galaxies implies a massive star formation rate, which 
is equivalent to the ccSN rate, of roughly $R_{SN}=0.014$\,year$^{-1}$. 
$R_{SN}=0.15$\,year$^{-1}$ implies that the expected number of candidates in the targeted galaxies should be $\simeq6$
($R_{SN} \times f_\eta \times t_d$ for $f_\eta \gtrsim 0.2$ and $t_d = 200$\,years) with a $>3\sigma$ chance of finding at least 1.
On the other hand, $R_{SN}=0.014$\,year$^{-1}$ reduces the probability to only about 40\% 
and implies that we need to study galaxies with an integrated star formation rate of 
20\,$M_\odot$\,year$^{-1}$ (10 times greater than what we have now) to have a $>3\sigma$ chance of 
finding at least 1 massive dust obscured star. 

In either case, our survey can be easily expanded to at least 10
times as many galaxies (and integrated star formation rate) simply 
using archival data from the SINGS, LVL, and $S^4G$ surveys, which
then probes rates far below those necessary to explain the
Galactic sources. In such an expanded survey, additional means of suppressing
contamination are important. The simplest method is to 
use the time variability of the mid-IR emission, since 
expanding shells of ejecta will also show a well defined
pattern of fading (see \citealp{ref:Kochanek_2012a}) in the warm 
\textit{Spitzer} bands (3.6 and 4.5\,$\micron$) and new \textit{Spitzer} 
observations would provide a time baseline of 5--10 years to 
search for such changes. Since the principle background
in our present survey appears to be extragalactic, time
variability is a powerful means of suppressing it. Galaxies
are not variable, and the mid-IR variability of quasars 
is both relatively weak and stochastic, with a structure
function of roughly $0.1(t/4~\hbox{year})^{1/2}$~mag
\citep{ref:Kozlowski_2010}. Two epochs separated by 6--12
months would further help to separate source classes
by constraining variability on shorter time scales.

\acknowledgments
We thank John Beacom for numerous helpful disucssions, and Jose Prieto, 
Todd Thompson, Shunsaku Horiuchi and Joe Antognini for helpful comments.
We extend our gratitude to the SINGS Legacy Survey and LVL Survey for making 
their data publicly available. This research has made use of NED, which is 
operated by the JPL and Caltech, under contract with NASA and the HEASARC 
Online Service, provided by NASA's GSFC. 
RK and KZS are supported in part by NSF grant AST-1108687. KZS and CSK are 
supported in part by NSF grant AST-0908816. 
\bibliographystyle{apj}
\bibliography{bibliography}
\clearpage

\clearpage

\clearpage

\clearpage

\clearpage

\clearpage

\clearpage

\clearpage

\clearpage

\clearpage

\end{document}